\documentclass[12pt]{article}
\usepackage{amsfonts,amsmath,amssymb,epsf,framed}
\usepackage{amsmath,braket}
\usepackage{graphicx,hyperref,color,comment}
\usepackage{cite}

\usepackage{subfigure}

\edef\restoreparindent{\parindent=\the\parindent\relax}
\usepackage{parskip}
\restoreparindent

\hypersetup{
    bookmarks=true,         
    unicode=false,          
    pdftoolbar=true,        
    pdfmenubar=true,        
    linktocpage=true,       
    pdffitwindow=false,     
    pdfstartview={FitH},    
    pdfnewwindow=true,      
    colorlinks=true,       
    linkcolor=blue,          
    citecolor=blue,        
    filecolor=blue,      
    urlcolor=blue           
}
\numberwithin{equation}{section}									

\let\a=\alpha \let\b=\beta  \let\d=\delta   \let\g=\gamma   \let\l=\lambda \let\m=\mu \let\n=\nu
  \let\r=\rho   \let\th=\theta     \let\z=\zeta
   \let\L=\Lambda   \let\S=\Sigma    
\def\nn{\nonumber}
\def\inf{\infty}

\def\pa{\partial}

\def\bs{\boldsymbol}

\begin{document}

\begin{titlepage}
\thispagestyle{empty}

\vspace*{-2cm}
\begin{flushright}
RUP-24-4\\
\vspace{2.0cm}
\end{flushright}

\bigskip

\begin{center}
\noindent{{\Large \textbf{Gauge Symmetries and Conserved Currents \\[6pt]in AdS/BCFT}}}\\
\vspace{2cm}

Kenta Suzuki
\vspace{1cm}\\

{\it Department of Physics, Rikkyo University, Toshima, Tokyo 171-8501, Japan}
\vspace{1mm}

\bigskip \bigskip
\vskip 3em
\end{center}

\begin{abstract}
In this paper, we study massless/massive vector and $p$-form field perturbations in AdS spacetime with an end-of-the-world brane.
By imposing $U(1)$ preserving Neumann boundary condition on the end-of-the-world brane, we study their spectrum and discuss their implications for dual BCFT operators.
When the perturbation is massless, the dual BCFT operator is a conserved current and we show that such an operator indeed satisfies the $U(1)$ preserving conformal boundary condition.
On the other hand, when the perturbation is massive, in general there exists non-vanishing perpendicular components of the dual BCFT operator, even in the massless limit.
We explain this difference between massless and massive perturbations from the point of view of the bulk gauge symmetry,
or equivalently from different structure of equations of motion.
We also find several brane-tension-independent modes in massless perturbations, and these are understood as boundary-condition-independent modes from the dual BCFT point of view.

\end{abstract}

\end{titlepage}

\newpage

\tableofcontents

\section{Introduction}
\label{sec:introduction}

\begin{figure}[t!]
	\begin{center}
        \vspace*{-30pt}
		\scalebox{1.2}{\includegraphics{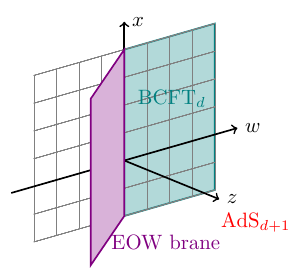}} 
        \vspace*{-4pt}
	\caption{A sketch of the AdS/BCFT setup. The end-of-the-world (EOW) brane is depicted in purple, and the asymptotic boundary is depicted in green, where the dual BCFT lives on. The boundary of the BCFT is a straight plane at $w=0$. The gravity dual of the BCFT lives in AdS$_{d+1}$ between the EOW brane and the asymptotic boundary.}
	\label{fig:adsbcft}
	\end{center}
\end{figure}

A boundary conformal field theory (BCFT) is a conformal field theory (CFT) defined on a manifold with codimension-one boundaries,
such that a maximal subgroup of conformal symmetry is preserved on the boundaries \cite{Cardy:1984bb, Cardy:2004hm, McAvity:1993ue, McAvity:1995zd}.
A holographic description of BCFTs was proposed in \cite{Takayanagi:2011zk, Fujita:2011fp},
inspired by the earlier works on the brane-world holography \cite{Randall:1999ee,Randall:1999vf,Karch:2000ct,Karch:2000gx}.
This duality is called AdS/BCFT correspondence, and its idea is to simply extend the boundary of BCFT towards inside of the bulk AdS spacetime,
which shapes the end-of-the-world (EOW) brane. (See figure~\ref{fig:adsbcft} for a sketch of the AdS/BCFT setup.)
The EOW brane has tension $T$, which is related to the brane location $\r_*$ by imposing Neumann boundary condition on the brane.
This AdS/BCFT correspondence recently found an interesting application for the black hole information paradox and derivation of the Page curve \cite{Almheiri:2019hni}.
In the context of AdS/BCFT correspondence, holographic stress-energy tensor has been investigated widely \cite{Nozaki:2012qd, Ugajin:2013xxa, Miao:2017gyt, Chu:2017aab, Seminara:2017hhh, Shimaji:2018czt, Caputa:2019avh, Hernandez:2020nem, Chalabi:2021jud, Kawamoto:2022etl, Suzuki:2022yru, Izumi:2022opi}.
On the other hand, holographic conserved currents associated to bulk gauge fields have not been discussed so extensively.
The purpose of this paper is to further investigate holographic conserved currents in the context of AdS/BCFT correspondence.

In addition to the maximal subgroup of conformal symmetry, in this paper, we consider a $U(1)$ preserving boundary condition \cite{Gaberdiel:2001zq,Green:1995ga,Recknagel:1998ih}.
From the point of view of BCFT, this means that the conserved current $J^\m$ associated to this $U(1)$ symmetry satisfies 
	\begin{align}
		J^w(w=0) \, = \, 0 \, , 
	\label{J_int}
    \end{align}
on the boundary. Here $w$ is the perpendicular direction to the boundary and we placed the boundary at $w=0$.
From the bulk point of view of AdS/BCFT, this condition implies that there is no coupling between bulk fields and brane-localized fields.
Gauge fields confined on the EOW brane were previously studied in \cite{Chamblin:2000ra, Cui:2023gtf}.
On the other hand, we study propagating bulk gauge fields, as well as massive fields, without brane-localized fields.
Our set-up is similar to the one considered in \cite{Kaloper:2000xa, Lu:2000xc, Oda:2000xh} in the context of brane world holography.
In contrast to these conserved currents $J^\m$, other vectorial operators $\mathcal{O}^\m$, in general can have a non-vanishing perpendicular component on the boundary:
	\begin{align}
		\mathcal{O}^w(w=0) \, \ne \, 0 \, .
	\label{O_int}
    \end{align}
These two types of operators $J^\m$ and $\mathcal{O}^\m$ in BCFT are associated to massless and massive vector fields in AdS spacetime through the AdS/BCFT correspondence.
The goal of this paper is to clarify the origin of these two different behaviors (\ref{J_int}) and (\ref{O_int}), from the bulk point of view of AdS/BCFT.
We will explain that the origin of these different behaviors are the presence or absence of {\it the bulk gauge symmetry, but not the value of masses}.
Therefore, even in the massless limit of a massive vector perturbation, in general one can obtain a non-vanishing perpendicular component on the boundary (\ref{O_int}).
We will also discuss $p$-form generalization of the above discussion.

Some other motivations of this work are as follows.
An abelian $p$-form gauge field in $d=2(p+1)$ dimensional flat space with a planar codimension-one boundary,
without any boundary interaction, is known to be BCFT \cite{Herzog:2017xha}.
Our discussion on the $p$-form gauge field perturbations in AdS would be useful when constructing a holographic model for such $p$-form BCFT.
Furthermore, when $p=1$ we can also induce boundary interactions.
One particular such model is known as boundary QED, where a 4-dimensional bulk Maxwell field and 3-dimensional planar boundary fermions interact on the boundary \cite{Herzog:2017xha, DiPietro:2019hqe, Bartlett-Tisdall:2023ghh}. 
This boundary QED was argued to model a time parity odd dynamics in a planar sheet of graphene in the continuum limit \cite{Dudal:2018mms}.
Even though we do not include boundary interactions,
our discussion on the $U(1)$ gauge field perturbation in AdS would be a first step toward constructing a holographic model for boundary (large $N$) QED.
It also seems like that a string theory embedding of AdS/BCFT correspondence requires some type of $p$-form gauge field \cite{Chiodaroli:2011nr,Chiodaroli:2012vc,Raamsdonk:2020tin,Coccia:2021lpp,Sugimoto:2023oul}.
Therefore, our study on the $p$-form perturbations might be useful to construct some string theory embedding of AdS/BCFT correspondence.

The rest of the paper is organized as follows.
In section~\ref{sec:background}, we introduce the background geometry of the AdS/BCFT setup and define our notations.
The following sections (from section~\ref{sec:massless-vector} to section~\ref{sec:massive p-form}),
we discuss perturbations on top of this background geometry and their implications for the dual BCFT operators.  
In section~\ref{sec:massless-vector}, we study a massless vector perturbation, where the equation of motion can be decomposed into each component, by using Lorentz gauge.
Then, the equation of motion becomes a second order homogeneous differential equation for each component, which leads to the behavior (\ref{J_int}).
In section~\ref{sec:massive-vector}, we study a massive vector perturbation, where one {\it cannot} decompose the equation of motion for each component, due to the absence of gauge symmetry. Then, the equation of motion becomes a second order {\it inhomogeneous} differential equation.
The general solution for the corresponding homogeneous equation is analogous to the massless solution, which leads to the behavior (\ref{J_int}) in the massless limit.
However a special solution for the inhomogeneous equation leads to the behavior (\ref{O_int}), even in the massless limit.
In section~\ref{sec:massless p-form} and section~\ref{sec:massive p-form}, we discuss $p$-form generalization of the above discussion,
and readers who are interested only in the vector cases can skip these sections, since the results are simple generalization of the vector cases.
On the other hand, readers who want to start from the most general discussions can skip vector sections and jump into the $p$-form sections.
Even though the vector cases can be obtained from the $p$-form cases by setting $p=1$, since vector fields are more ubiquitous, we decided to present them separately.
In section~\ref{sec:conclusions}, we conclude this paper, and the appendix gives the formulae to derive the dual BCFT operators from asymptotics of bulk fields.

\section{Background}
\label{sec:background}
In this section, we summarize the background geometry of the AdS/BCFT setup \cite{Takayanagi:2011zk, Fujita:2011fp}.

For the background, we have the total action
	\begin{align}
		I_{\textrm{background}} \, = \, I_{\textrm{bulk}} \, + \, I_{\textrm{brane}} \, + \, I_{\textrm{bdy}} \, ,
	\end{align}
where 
	\begin{align}
		I_{\textrm{bulk}} \, &= \, - \frac{1}{16\pi G_N} \int_M d^{d+1}x \sqrt{g} \, (R - 2 \L) \, , \\
		I_{\textrm{brane}} \, &= \, - \frac{1}{8\pi G_N} \int_Q d^dx \sqrt{h} \, (K_Q - T) \, , \label{I_brane} \\
		I_{\textrm{bdy}} \, &= \, - \frac{1}{8\pi G_N} \int_\S d^dx \sqrt{\g} \, K_\g  \, .
	\end{align}
We denote the bulk AdS$_{d+1}$ spacetime by $M$, which is cut by the EOW brane $Q$. The asymptotic AdS boundary is denoted by $\S$, where the dual BCFT lives on.
The cosmological constant is given by
	\begin{align}
		\L \, = \, - \frac{d(d-1)}{2\, \ell_{\textrm{AdS}}^2} \, , 
	\end{align}
and hereafter we set $\ell_{\textrm{AdS}}=1$. $h$ and $\g$ are induced metrics on the EOW brane and the asymptotic AdS boundary, respectively.
$K$'s are trace of the extrinsic curvatures and $T$ is the brane tension.

Since the bulk action is just the Einstein-Hilbert action with negative cosmological constant, the solution is given by an empty AdS$_{d+1}$.
For AdS$_{d+1}$, the Poincare metric is written as
	\begin{align}
		ds_{d+1}^2 \, = \, \frac{dz^2+dw^2 + \sum_{i=1}^{d-2} dx_i^2}{z^2} \, .
	\label{Poincare}
	\end{align}
However, these exsit more convenient coordinates for a study of AdS/BCFT. By coordinate transformations
	\begin{align}
		z \, = \, \frac{y}{\cosh \r} \, , \qquad w \, = \, y \tanh \r \, ,
	\label{coord-transf}
	\end{align}
we can move to the hyperbolic slicing coordinates 
	\begin{align}
		ds_{d+1}^2 \, = \, d\r^2 \, + \, \cosh^2 \r \left( \frac{dy^2 + \sum_{i=1}^{d-1} dx_i^2}{y^2} \right) \, .
	\end{align}
In the following we denote this metric by 
	\begin{align}
		ds_{d+1}^2 \, = \, d\r^2 + a(\r)^2 \g_{\m\n} dq^\m dq^\n \, ,
	\label{metric}
	\end{align}
such that 
	\begin{align}
		a(\r) \, := \, \cosh \r \, ,
	\end{align}
$q^\m=\{x^i, y\}$ and $\g_{\m\n}$ is the $d$-dimensional Poincare metric.
In this paper, we will mostly follow the notation of \cite{Izumi:2022opi}.
In particular, the indices $\m, \n, \l, \cdots$ denote the $d$-dimensional coordinates without $\r$-direction.
We use capital indices $M, N, L, \cdots $ to denote ($d+1$) dimensional coordinates and small indices $i, j, k, \cdots $ to denote ($d-1$) dimensional coordinates. 

We impose Neumann boundary condition on the EOW brane, which is explicitly written as
	\begin{align}
		K_{\m\n} \, = \, (K - T) h_{\m\n} \, , 
	\end{align}
where the $K_{\m\n}$ (and $K$) is the EOW brane extrinsic curvature (and its trace).
This condition fixes the brane location.
We place the EOW brane at $\r=\r_*$, and this position of the brane is determined by the boundary condition as \cite{Takayanagi:2011zk, Fujita:2011fp}
	\begin{align}
		T \, = \, (d-1) \tanh \r_* \, .
	\label{tension}
    \end{align}

\section{Massless Vector Perturbations}
\label{sec:massless-vector}
Now we study vectorial perturbations on top of the background discussed in section~\ref{sec:background}.
In this section, we study massless $U(1)$ gauge field perturbations. We will also discuss massive vector case in section~\ref{sec:massive-vector}.
Massless $U(1)$ gauge field in empty AdS was previously discussed in \cite{Allen:1985wd, DHoker:1998bqu, DHoker:1999bve}.

We induce only $d$-dimensional perturbations and we impose $A^\r=0$.
Therefore, the perturbation action is given by
	\begin{align}
		I_{\textrm{EM}} \, = \, - \frac{1}{4} \int_M d^{d+1}x \sqrt{g} \, F^{\m\n} F_{\m\n} \, .
	\label{I_EM}
	\end{align}
The equation of motion derived from (\ref{I_EM}) is 
	\begin{align}
		\nabla_M F^{MN} \, = \, 0\, ,
	\end{align}
where $\nabla_M$ is the covariant derivative with respect to the background (\ref{metric}) and the field strength is $F^{MN} = \nabla^M A^N - \nabla^N A^M$.
We now also impose the Lorentz gauge for the $d$-dimensional coordinates
	\begin{align}
		\tilde{\nabla}_\m A^\m \, = \, 0 \, ,
	\label{Lorentz gauge}
	\end{align}
where $\tilde{\nabla}_\m$ is the covariant derivative with respect to the $d$-dimensional Poincare metric $\g_{\m\n}$.
We note that with the condition $A^\r =0$, this $d$-dimensional Lorentz gauge also implies ($d+1$)-dimensional Lorentz gauge
	\begin{align}
		\nabla_M A^M \, = \, 0 \, .
	\end{align}
Upon this Lorentz gauge, the equation of motion is rewritten as \cite{Allen:1985wd}
	\begin{align}
		(\Box + d) A^M \, = \, 0 \, ,
	\end{align}
where $\Box:=g^{MN} \nabla_M \nabla_N$.
Now we decompose the $\r$-direction. Then the equation of motion leads to
	\begin{align}
		\left[ a^{-2} (\tilde{\Box} + 1 ) + \pa_\r^2 + (d+2) \tanh \r \, \pa_\r + 2d - \frac{d}{\cosh^2 \r} \right] A^\n \, = \, 0 \, , 
	\end{align}
with $\tilde{\Box} := \g^{\m\n} \tilde{\nabla}_\m \tilde{\nabla}_\n$.
We assume the separable form of the solutions
	\begin{align}
		A^\m(\r, x, y) \, = \, a(\r)^{-1} \sum_{n=0}^\inf \, \mathcal{R}_n(\r) Y^\m_n(x, y) \, .
	\end{align}
Here, we separated a factor of $a(\r)^{-1}$ in order to follow the convention of \cite{Izumi:2022opi}.
Then, the equation of motion is decomposed into the $\r$-direction and the remaining $d$-dimensional equations as 
	\begin{gather}
		\mathcal{R}_n'' \, + \, d \, \tanh \r \, \mathcal{R}_n' \, + \, (d-1) \mathcal{R}_n \, = \, \frac{-\l_n}{\cosh^2 \r} \, \mathcal{R}_n \, , \label{eq_rho} \\
		(\tilde{\Box} + 1 ) Y_n^\m \, = \, \l_n Y_n^\m \, , \label{eq_x} 
	\end{gather}
where the prime denotes a derivative with respect to $\r$.

\subsection{$\r$-direction}
\label{sec:rho}
The $\r$-direction equation (\ref{eq_rho}) is solved by the associated Legendre functions as
	\begin{align}
		\mathcal{R}_n(\r) \, = \, (1-\z^2)^{\frac{d}{4}} \left[ c_1 \, P_{\n-\frac{1}{2}}^\m(\z) \, + \, c_2 \, Q_{\n-\frac{1}{2}}^\m(\z) \right] \, ,
	\label{R_n}
	\end{align}
where 
	\begin{align}
		\z \, = \, \tanh \r \, , \qquad
		\m \, = \, \frac{d-2}{2} \, , \qquad
		\n \, = \, \frac{1}{2} \sqrt{(d-1)^2 + 4\l_n} \, .
	\end{align}

We note that for a massive scalar field, we have the same expression for the $\r$-direction wavefunction as discussed in \cite{Izumi:2022opi},
except that $\m$ is given by
	\begin{align}
		\m \, = \, \frac{1}{2} \, \sqrt{d^2 + 4m^2} \, ,
	\end{align}
where $m$ is the mass of the scalar field.
This implies that the massless gauge field wavefunction is related to that of massive scalar field with mass square 
	\begin{align}
		m^2 \, = \, - (d-1) \, .
	\end{align}
The same relation between the massless gauge field and a massive scalar field was also noticed in empty AdS in \cite{DHoker:1998bqu}.

For the asymptotic AdS boundary ($\z \to 1$ or $\r \to \inf$), we impose the standard Dirichlet boundary condition.
This fixes the relation between the two coefficients $c_1$ and $c_2$ in (\ref{R_n}).
After imposing this boundary condition, the solution is written as \cite{Izumi:2022opi}
	\begin{align}
		\mathcal{R}_n(\r) \, = \, B_0 \, (1-\z^2)^{\frac{d}{4}} \, P_{\n-\frac{1}{2}}^{-\m}(\z) \, ,
	\end{align}
where $B_0$ is a new numerical coefficient and we note that the order of the associated Legendre function is now replaced by $-\m$.
This solution behaves as 
	\begin{align}
		\mathcal{R}_n(\r) \, \sim \, (1- \z)^{\frac{d-1}{2}} \, ,
	\end{align}
towards the asymptotic AdS boundary ($\z \to 1$).
Therefore, the behavior of the original gauge field is
	\begin{align}
		A^\m(\r, x) \, \sim \, e^{-d \r} \, .
	\label{A-asympto}
	\end{align}

On the EOW brane, we impose Neumann boundary condition \cite{Takayanagi:2011zk, Fujita:2011fp}
	\begin{align}
		0 \, = \, F_{\r \m} \big|_{\r = \r_*} \, .
	\end{align}
Since we set $A^\r=0$, this condition implies 
	\begin{align}
		0 \, = \, \pa_\r \big( a^2 A^\m \big) \big|_{\r = \r_*} \, ,
	\end{align}
or equivalently
	\begin{align}
		0 \, = \, \frac{d}{d\z} \big( a \, \mathcal{R}_n \big) \big|_{\z=\z_*} \, . 
	\end{align}
where $\z_*=\tanh \r_*$.
Using a property of the Legendre function, this condition can be written in a simply form as in the gravity case \cite{Izumi:2022opi}:
	\begin{align}
		0 \, = \, P_{\n-\frac{1}{2}}^{1-\m}(\z_*) \, . 
	\label{BCs}
	\end{align}
This condition quantizes the eigenvalue (or Kaluza-Klein mass) $\l_n$.
Obtaining exact analytical solutions for this condition seems difficult, except $d=3$ and $d=5$ cases.
For $d=3$, the Neumann boundary condition is written as
	\begin{align}
		0 \, = \, P_{\n-\frac{1}{2}}^{\frac{1}{2}}(\z_*) \, = \, \sqrt{\frac{2}{\pi \sin \th_*}} \, \cos(\n \th_*) \, ,
	\end{align}
where we defined $\th_*$ by 
	\begin{align}
		\cos \th_* = \z_*=\tanh \r_* \, .
	\end{align}
This condition is solved by 
	\begin{align}
		\n_n \, = \, \left( \frac{2n+1}{2} \right) \frac{\pi}{\th_*} \, , \qquad (n=0,1,2, \cdots )
	\end{align}
For $d=5$, the Neumann boundary condition is written as
	\begin{align}
		0 \, = \, P_{\n-\frac{1}{2}}^{-\frac{1}{2}}(\z_*) \, = \, \sqrt{\frac{2}{\pi \sin \th_*}} \, \frac{\sin(\n \th_*)}{\n} \, ,
	\end{align}
This was already discussed in \cite{Izumi:2022opi} and the solutions are given by
	\begin{align}
		\n_n \, = \, \frac{(n+1)\pi}{\th_*} \, , \qquad (n=0,1,2, \cdots )
	\end{align}

\begin{figure}[t!]
	\begin{center}
		\scalebox{0.71}{\includegraphics{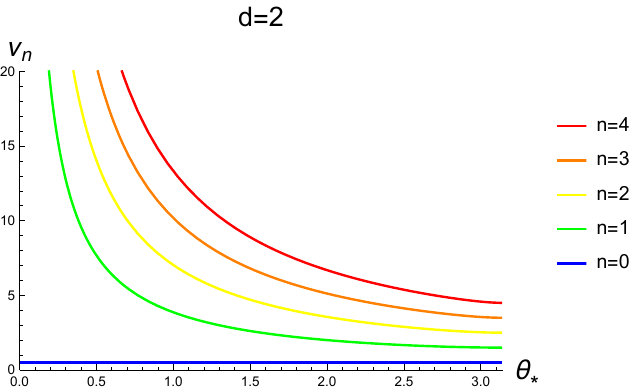}} \ \quad \scalebox{0.71}{\includegraphics{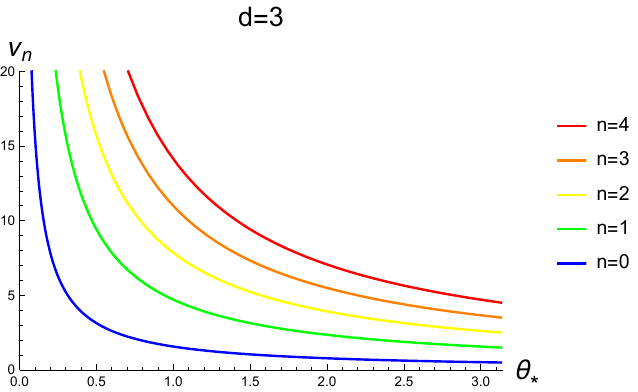}} \\[8pt]
		\scalebox{0.71}{\includegraphics{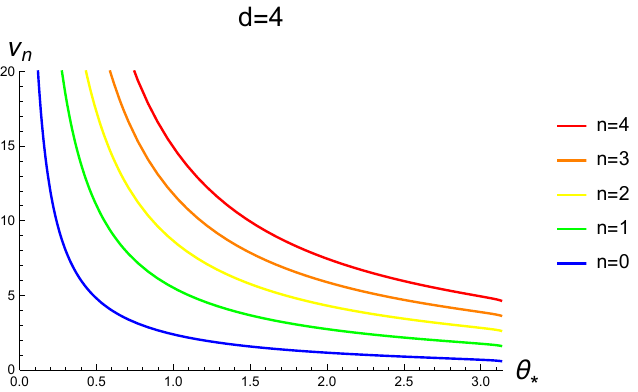}} \ \quad \scalebox{0.71}{\includegraphics{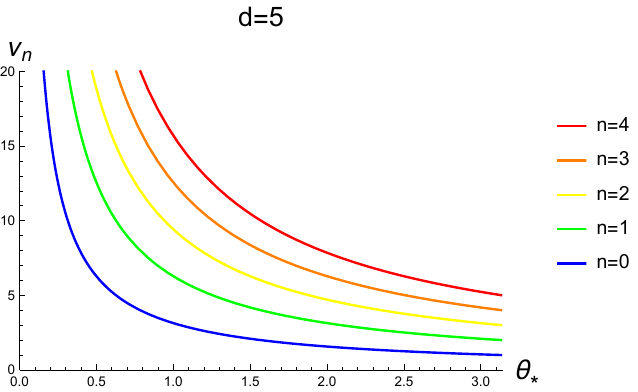}} 	
	\caption{Brane tension dependence of numerical solutions for the lowest five modes in AdS$_{d+1}$ with $d=2,3,4$ and $5$.}
	\label{fig:rho*_dependence}
	\end{center}
\end{figure}

In figure~\ref{fig:rho*_dependence}, we show numerical solutions for low lying modes $\n_n$ with $n=0,1,2,3,4$.
We notice that for $d=2$, the lowest mode ($n=0$) is actually independent of the brane location $\th_*$, or the brane tension $T$ through the relation (\ref{tension}).
This can be explained as follows.
For $d=2$, the Neumann boundary condition (\ref{BCs}) can be written as
	\begin{align}
		0 \, = \, P_{\n-\frac{1}{2}}^{1}(\z_*) \, = \, - (1- \z_*^2)^{\frac{1}{2}} \left[ \frac{d}{d\z} \, P_{\n-\frac{1}{2}}(\z) \right] \bigg|_{\z=\z_*}\, . 
	\label{d=2_BC}
    \end{align}
Therefore, when the Legendre function $P_{\n-\frac{1}{2}}(\z)$ is constant (i.e. independent of $\z$), such $\n$ is a solution.
This happens when $\n=1/2$, since $P_0(x)=1$.
Therefore for $d=2$, the lowest mode is given by
	\begin{align}
		\qquad \n_0 \, = \, \frac{1}{2} \, , \qquad (\textrm{for} \ \, d=2)
	\end{align}
It would be interesting to further explore physical meaning of this constant mode for the dual BCFT$_2$, but we leave this question for a future work.

For the asymptotic spectrum, we can also find an analytical expression. (For the scalar perturbation, this was found in \cite{Reeves:2021sab}.)
For large $\n$ (with fixed $\m$), the boundary condition (\ref{BCs}) can be approximated by
	\begin{align}
		0 \, &= \, \frac{1}{\n^{\m-1}} \left( \frac{\th_*}{\sin \th_*} \right)^{\frac{1}{2}} \Big( J_{\m -1}(\n \th_*) + \mathcal{O}(\n^{-1}) \Big) \nn\\
        & \approx \, \frac{1}{\n^{\m-1}} \left( \frac{2}{\pi \n \sin \th_*} \right)^{\frac{1}{2}} \cos\left( \n \th_* - \frac{\pi \m}{2} - \frac{\pi}{4} \right) \, .
	\end{align}
Therefore, the asymptotic spectrum is found as
	\begin{align}
		\n_n \, \approx \, \frac{\pi}{2\th_*} \left( \m + \frac{3}{2} + 2n \right) \, , \qquad (\textrm{for\ \, large }\,  n)
	\label{asymp_spec}
    \end{align}

\subsection{$x$-direction}
\label{sec:x}
Now we consider the remaining $d$-dimensional equation (\ref{eq_x}).
We study this equation on the $d$-dimensional Poincare metric 
	\begin{align}
		ds_d^2 \, = \, \g_{\m\n} dq^\m dq^\n \, = \, \frac{dy^2 + \sum_{i=1}^{d-1} dx_i^2}{y^2} \, .
	\end{align}
First we note that the Lorentz gauge (\ref{Lorentz gauge}) is explicitly written as
	\begin{align}
		\pa_i Y^i \, = \, - \pa_y Y^y \, + \, \frac{d}{y} \, Y^y \, .
	\label{Lorentz gauge2}
	\end{align}
Using this condition, we can write the $x$ -direction equation (\ref{eq_x}) explicitly as
	\begin{align}
		0 \, &= \, \Box_0 Y^y - \, \left( \frac{d+2}{y} \right) \pa_y Y^y + \, \left( \frac{2d+2-\l_n}{y^2} \right) Y^y \, , \label{eq_Y^y} \\
		0 \, &= \, \Box_0 Y^i - \, \frac{d}{y} \, \pa_y Y^i - \, \frac{2}{y} \, \pa_i Y^y + \, \left( \frac{d-\l_n}{y^2} \right) Y^i \, , \label{eq_Y^i} 
	\end{align}
where
	\begin{align}
		\Box_0 \, := \, \pa_y^2 + \pa_i \pa_i \, .
	\end{align}

The $Y^y$ equation (\ref{eq_Y^y}) is solved as
	\begin{align}
		Y^y(x, y) \, = \, y^{\frac{d+3}{2}} \left[ c_3 \, J_\n(\sqrt{-\bs{k}^2} y) \, + \, c_4 \, Y_\n(\sqrt{-\bs{k}^2} y) \right] e^{i \bs{k} \cdot \bs{x}} \, .
	\label{Y^y}
	\end{align}
We impose Dirichlet boundary condition at $y=0$ as in \cite{Izumi:2022opi}, which sets $c_4=0$.
Therefore, the asymptotic behavior ($y \to 0$) is given by
	\begin{align}
		Y^y(x, y) \, \sim \, y^{\frac{d+3}{2}+\n} \, .
	\end{align}

For the remaining modes $Y^i$, we decompose it into helicity-$0$ component $\phi$ and helicity-$1$ component $V^i$ as
	\begin{align}
		Y^i \, = \, \pa^i \phi \, + \, V^i \, ,
	\end{align}
where 
	\begin{align}
		\pa_i V^i \, = \, 0 \, .
	\end{align}
The helicity-$0$ component $\phi$ is fixed by the Lorentz gauge (\ref{Lorentz gauge2}) as
	\begin{align}
		\phi \, = \, \frac{1}{\bs{k}^2} \left( \pa_y Y^y \, - \, \frac{d}{y} \, Y^y \right) \, .
	\end{align}
Therefore, the asymptotic behavior of the helicity-$0$ component $Y_{(0)}^i=\pa^i \phi $ is given by
	\begin{align}
		Y_{(0)}^i(x, y) \, \sim \, y^{\frac{d+1}{2}+\n} \, .
	\end{align}
For the helicity-$1$ component $Y_{(1)}^i=V^i$, $Y^y$ does not contribute. Therefore, the equation (\ref{eq_Y^i}) is simply solved as
	\begin{align}
		Y_{(1)}^i(x, y) \, = \, y^{\frac{d+1}{2}} \left[ c_5 \, J_\n(\sqrt{-\bs{k}^2} y) \, + \, c_6 \, Y_\n(\sqrt{-\bs{k}^2} y) \right] e^{i \bs{k} \cdot \bs{x}} E^i(\bs{k}, s) \, ,
	\label{Y^i}
	\end{align}
where $E^i(\bs{k}, s)$ is a basis vector satisfying 
	\begin{align}
	k_i E^i(\bs{k}, s) \, = \, 0 \, , \qquad E_i(\bs{k}, s) E^i(\bs{k}, s') \, = \, \d_{ss'} \, .
	\end{align}
Here, $s$ is the label representing different choices of basis vectors.
Therefore, the asymptotic behavior of the helicity-$1$ component is same as that of helicity-$0$ component, so the parallel components have
	\begin{align}
		Y^i(x, y) \, \sim \, y^{\frac{d+1}{2}+\n} \, .
	\end{align}

\subsection{Holographic Conserved Currents}
\label{sec:currents}
The holographic conserved currents $J^\m$ are obtained from bulk gauge field perturbations by 
	\begin{align}
		J^\m(x, w) \, = \, \lim_{z \to 0} \, z^{-d} A^\m(x, w, z) \, ,
	\label{J^a(1)}
	\end{align}
in the Poincare coordinates. This formula is derived in Appendix~\ref{app:scaling}.
Using the coordinate transformations (\ref{coord-transf}), we can rewrite this in the hyperbolic slicing coordinates as
	\begin{align}
		J^\m(x, w) \, = \, \lim_{\r \to \inf} \left[ \left( \frac{e^\r}{2y}\right)^d A^\m(\r, x, y) \right] \Bigg|_{y=w}\, .
	\label{J^a(2)}
	\end{align}
Since the asymptotic behavior of the gauge field is given by (\ref{A-asympto})
	\begin{align}
		A^\m(\r, x, y) \, \sim \, e^{-d \r} \, ,
	\end{align}
the $\r \to \inf$ limit in (\ref{J^a(2)}) indeed gives a finite value.
Finally combining with the asymptotic behavior of $Y^\m$ discussed in section~\ref{sec:x}, near-the-boundary behavior ($w \to 0$) of the holographic conserved currents are found as
	\begin{align}
		J^w(x, w) \, \propto \, w^{\, \n+\frac{-d+3}{2}} \, , \qquad J^i(x, w) \, \propto \, w^{\, \n+\frac{-d+1}{2}} \, .
	\end{align}
Since $\l_n$ is non-negative, we have
	\begin{align}
		\n \, \ge \, \frac{d-1}{2} \, .
	\end{align}
This implies that at most the powers of the perpendicular direction $w$ are given by
	\begin{align}
		J^w(x, w) \, \propto \, w \, , \qquad J^i(x, w) \, \propto \, w^{\, 0} \, = \, 1 \, .
	\end{align}
Therefore, the perpendicular component of the holographic conserved current $J^w$ vanishes at the boundary $J^w(w=0)=0$,
while the transverse components can have finite non-vanishing values at the boundary $w=0$.

\section{Massive Vector Perturbations}
\label{sec:massive-vector}
In this section, we study massive vector perturbations on top of the background discussed in section~\ref{sec:background}.
Massive vector fields in empty AdS were previously discussed in \cite{Allen:1985wd, Mueck:1998iz, lYi:1998akg}.

Unlike the massless case, we cannot impose the gauge condition $A^\r=0$, since there is no gauge symmetry in the present case.
Therefore, we induce the full $(d+1)$-dimensional vector perturbations and the perturbation action is given by
	\begin{align}
		I_{\textrm{v}} \, = \, - \int_M d^{d+1}x \sqrt{g} \left[ \, \frac{1}{4} \, F^{MN} F_{MN} + \frac{m^2}{2} A^M A_M \right] \, .
	\end{align}
The equation of motion derived from this action is 
	\begin{align}
		\nabla_M F^{MN} \, = \, m^2 A^N \, .
	\end{align}
Even though we do not have any gauge fixing condition, due to the anti-symmetric nature of $F^{MN}$, we get a supplemental condition \cite{Allen:1985wd}
	\begin{align}
		\nabla_M A^M \, = \, 0 \, ,
	\label{sup-cond}
	\end{align}
from the equation of motion.
Using this supplemental condition, the equation of motion is rewritten as
	\begin{align}
		(\Box + d - m^2) A^M \, = \, 0 \, .
	\label{massive-eq}
	\end{align}

As in the massless case, once we separate the $\r$-direction, the equations of motion read
	\begin{align}
		\left[ a^{-2} \, \tilde{\Box} + \, \pa_\r^2 + (d+2) \tanh \r \, \pa_\r + 2d - m^2 - \frac{d}{\cosh^2 \r} \right] A^\r \, = \, 0 \, , 
	\label{eq-A^rho}
	\end{align}
and 
	\begin{align}
		\left[ a^{-2} (\tilde{\Box} + 1 ) + \pa_\r^2 + \, (d+2) \tanh \r \, \pa_\r + 2d - m^2 - \frac{d}{\cosh^2 \r} \right] A^\n \, = \, -2 \tanh \r \, \pa^\n A^\r \, .
	\label{eq-A^mu}
	\end{align}
To write these equations, we also used the supplemental condition (\ref{sup-cond}), whose $\r$-separated form is given by
	\begin{align}
		\tilde{\nabla}_\m A^\m \, = \, -\pa_\r A^\r \, - \, d \tanh \r \, A^\r \, .
	\end{align}
We note that for this massive vector case, the $M=\r$ component of the equation of motion leads to a non-trivial equation, which we need to solve.
Furthermore, a slight difficultly in the current massive vector case is that the second equation is an {\it inhomogeneous} differential equation and it has a source contribution from $A^\r$ in the right-hand side.
We note that this structural difference of the equations of motion is due to the absence of a bulk gauge symmetry to set $A^\r=0$.
As we indicated in the Introduction, this structural difference of the equations of motion eventually leads to the different behavior of the perpendicular component of the dual BCFT operator, as we will see below.

\subsection{Solution for $A^\r$}
\label{sec:A^rho}
We first solve the equation (\ref{eq-A^rho}) for $A^\r$.
As in the massless case, we assume the separable form of the solution and take an ansatz
	\begin{align}
		A^\r(\r, x, y) \, = \, a(\r)^{-1} \sum_{n=0}^\inf \, \mathcal{R}_n(\r) Y^\r_n(x, y) \, .
	\label{A^rho}
	\end{align}
Then, the equation of motion is decomposed as 
	\begin{gather}
		\mathcal{R}_n'' \, + \, d \, \tanh \r \, \mathcal{R}_n' \, + \, (d-1-m^2) \mathcal{R}_n \, = \, \frac{-\l_n}{\cosh^2 \r} \, \mathcal{R}_n \, , \label{eq_rho2} \\
		\tilde{\Box} \, Y_n^\r \, = \, \l_n Y_n^\r \, . \label{eq_x2} 
	\end{gather}

The equation for $\mathcal{R}_n$ is basically the same structure as in the massless case, so the solution is given by
	\begin{align}
		\mathcal{R}_n(\r) \, = \, B_0 \, (1-\z^2)^{\frac{d}{4}} \, P_{\n-\frac{1}{2}}^{-\m}(\z) \, ,
	\label{R_n-solution}
	\end{align}
but the order of the associated Legendre function is now modified by the mass as
	\begin{align}
		\z \, = \, \tanh \r \, , \qquad
		\m \, = \, \frac{1}{2} \sqrt{(d-2)^2 + 4m^2} \, , \qquad
		\n \, = \, \frac{1}{2} \sqrt{(d-1)^2 + 4\l_n} \, .
	\label{massive-mu}
	\end{align}
This solution behaves as 
	\begin{align}
		\mathcal{R}_n(\r) \, \sim \, (1- \z)^{\frac{d+2\m}{4}} \, ,
	\end{align}
towards the asymptotic AdS boundary ($\z \to 1$).
Therefore, the behavior of the original gauge field is
	\begin{align}
		A^\r(\r, x, y) \, \sim \, e^{- (\frac{d+2}{2} + \m ) \r} \, .
	\label{A-asympto2}
	\end{align}
Once we impose a boundary condition on the EOW brane at $\r = \r_*$, this condition quantizes the eigenvalue $\l_n$.

Next for the $Y_n^\r$, since it behaves as a scalar field in equation (\ref{eq_x2}), we don't need to worry about the rest of the component $A^\m$ to solve this equation.
It is explicitly written as
	\begin{align}
		0 \, = \, \Box_0 Y^\r - \, \left( \frac{d-2}{y} \right) \pa_y Y^\r - \, \frac{\l_n}{y^2} \, Y^\r \, .
	\end{align}
Therefore, the solution is given by
	\begin{align}
		Y^\r(x, y) \, \propto \, y^{\frac{d-1}{2}} \, J_\n(\sqrt{-\bs{k}^2} y) \, e^{i \bs{k} \cdot \bs{x}} \, .
	\end{align}
The asymptotic behavior ($y \to 0$) of this solution is given by
	\begin{align}
		Y^\r(x, y) \, \sim \, y^{\frac{d-1}{2}+\n} \, .
	\label{Y^rho}
	\end{align}

\subsection{Solution for $A^\m$}
\label{sec:A^rho}
For the equation (\ref{eq-A^mu}) for $A^\m$, we first study the homogeneous equation.
For the homogeneous solution, we assume the separable form of the solution and take an ansatz
	\begin{align}
		A_{\textrm{hom}}^\m(\r, x, y) \, = \, a(\r)^{-1} \sum_{n=0}^\inf \, \mathcal{R}_n(\r) Y^\m_n(x, y) \, .
	\end{align}
Then, the equation of motion is decomposed as 
	\begin{gather}
		\mathcal{R}_n'' \, + \, d \, \tanh \r \, \mathcal{R}_n' \, + \, (d-1-m^2) \mathcal{R}_n \, = \, \frac{-\l_n}{\cosh^2 \r} \, \mathcal{R}_n \, , \\
		(\tilde{\Box} + 1 ) \, Y_n^\m \, = \, \l_n Y_n^\m \, . 
	\end{gather}
We can see that the equation for $\mathcal{R}_n$ is the same as $A^\r$ case discussed in the previous subsection.
Therefore, the solution is given by (\ref{R_n-solution}) with (\ref{massive-mu}).
The quantization condition of $\n_n$ is given by the same form as in (\ref{BCs}) but $\m$ is now replaced by (\ref{massive-mu}).
We note that the brane-tension-independent mode we found in the massless case in $d=2$ does not exist in the present massive case.
This is because for $m^2>0$ and $d\ge2$, we have $\m>0$.
Therefore for massive case, we cannot use the formula like (\ref{d=2_BC}) to express the boundary condition as a derivative with respect to $\z$.

On the other hand, the equation for $Y_n^\m$ is the same as massless case discussed in the previous section, provided that the eigenvalue $\l_n$ is appropriately quantized in the current massive case. 
Therefore, besides the precise quantization condition for $\l_n$, the solutions are given by (\ref{Y^y}) and (\ref{Y^i}) with $c_4=c_6=0$.
In particular, the asymptotic behavior ($y \to 0$) of the homogeneous solutions are given by
	\begin{align}
		Y_{\textrm{hom}}^y(x, y) \, \sim \, y^{\frac{d+3}{2}+\n} \, , \qquad Y_{\textrm{hom}}^i(x, y) \, \sim \, y^{\frac{d+1}{2}+\n} \, .
	\label{homo-sol}
	\end{align}

Next, we consider a special solution of the inhomogeneous equation (\ref{eq-A^mu}).
It looks difficult to find an exact solution, since the solution of $A^\r$ is given in the summation form as in (\ref{A^rho}).
Therefore, in this subsection, let us just determine the leading asymptotic behavior of a special solution.
For the $\r$ direction, we use the following ansatz for the special solution
	\begin{align}
		\hat{A}^\m(\r, x, y) \, = \, e^{- (\frac{d+2}{2} + \m ) \r} \, \hat{Y}^\m(x, y) \, .
	\end{align}
We use a hat to denote this special solution. With this ansatz, the leading contribution (in large $\r$) of the inhomogeneous equation (\ref{eq-A^mu}) identically vanishes and the subleading contribution is given by
	\begin{align}
		\big( \tilde{\Box} + 1 - \hat{\l} \big) \hat{Y}^\m \, = \, -2 \, \pa_\m Y^\r \, ,
	\end{align}
where we defined 
	\begin{align}
		\hat{\l} \, := \, d - \left( \frac{d+2}{2} \right) \left( \frac{d+2}{2} + \m \right) \, .
	\end{align}
With this definition of $\hat{\l}$, the LHS becomes the same form as (\ref{eq_x}), whose explicit form is given by (\ref{eq_Y^y}) and (\ref{eq_Y^i}).
Therefore for $\hat{Y}^\m$, we need to solve 
	\begin{align}
		&\Box_0 \hat{Y}^y - \, \left( \frac{d+2}{y} \right) \pa_y \hat{Y}^y + \, \left( \frac{2d+2-\hat{\l}}{y^2} \right) \hat{Y}^y \, = \, - \frac{2}{y^2} \, \pa_y Y^\r \, ,\\
		&\Box_0 \hat{Y}^i - \, \frac{d}{y} \, \pa_y \hat{Y}^i + \, \left( \frac{d-\hat{\l}}{y^2} \right) \hat{Y}^i \, = \, \frac{2}{y} \, \pa_i \hat{Y}^y - \, \frac{2}{y^2} \, \pa_i Y^\r \, .
	\end{align}
Comparing with the asymptotic behavior of $Y^\r$ (\ref{Y^rho}), we find the asymptotic behavior of the special solution as
	\begin{align}
		\hat{Y}^y \, \sim \, y^{\frac{d-3}{2}+\n} \, , \qquad \hat{Y}^i \, \sim \, y^{\frac{d-1}{2}+\n}  \, .
	\end{align}
Therefore for all components of $\hat{Y}^\m$, we have found that the exponents of the asymptotic behavior are smaller than those of the general solution for the homogeneous equation (\ref{homo-sol}).

\subsection{Dual Vector Operators}
The dual vector operator $\mathcal{O}^\m$ is obtained from bulk massive vector field perturbation by 
	\begin{align}
		\mathcal{O}^\m(x, w) \, = \, \lim_{z \to 0} \, z^{-\left( \frac{d+2}{2} + \m \right)} A^\m(x, w, z) \, ,
	\label{J^a(3)}
	\end{align}
in the Poincare coordinates. This formula is derived in Appendix~\ref{app:scaling}.
Using the coordinate transformations (\ref{coord-transf}), we can rewrite this in the hyperbolic slicing coordinates as
	\begin{align}
		\mathcal{O}^\m(x, w) \, = \, \lim_{\r \to \inf} \left[ \left( \frac{e^\r}{2y}\right)^{\frac{d+2}{2} + \m} A^\m(\r, x, y) \right] \Bigg|_{y=w}\, .
	\label{J^a(4)}
	\end{align}
Since the asymptotic behavior of the massive vector field is given by (\ref{A-asympto2})
	\begin{align}
		A^\m(\r, x, y) \, \sim \, e^{- (\frac{d+2}{2} + \m ) \r} \, ,
	\end{align}
for both the homogeneous and inhomogeneous solutions, the $\r \to \inf$ limit in (\ref{J^a(3)}) indeed gives a finite value.
Therefore, we next need to consider the asymptotic behavior of $Y^\m$.
Let us separately consider the contributions from the homogeneous and inhomogeneous solutions.
First, the contribution from the homogeneous solution is found as
	\begin{align}
		\mathcal{O}_{\textrm{hom}}^w(x, w) \, \propto \, w^{\, \n+\frac{1}{2}-\m} \, , \qquad \mathcal{O}_{\textrm{hom}}^i(x, w) \, \propto \, w^{\, \n-\frac{1}{2}-\m} \, .
	\end{align}
If we take the massless limit $m^2 \to 0$, these scaling behaviors coincide with those of the conserved currents, discussed in section \ref{sec:currents}.
Therefore, in the massless limit, $\mathcal{O}_{\textrm{hom}}^w$ vanishes on the boundary $w=0$.

Next we show that the contribution from the inhomogeneous solution gives qualitatively different behavior compared with the homogeneous solutions.
The contribution from the inhomogeneous solution is found as
	\begin{align}
		\mathcal{O}_{\textrm{inh}}^w(x, w) \, \propto \, w^{\, \n-\frac{5}{2}-\m} \, , \qquad \mathcal{O}_{\textrm{inh}}^i(x, w) \, \propto \, w^{\, \n-\frac{3}{2}-\m} \, .
	\end{align}
Even in the massless limit, $\mathcal{O}_{\textrm{inh}}^w$ can have a negative power in $w$ in the close to the boundary limit.
Therefore, in general the contribution from the inhomogeneous solution $\mathcal{O}_{\textrm{inh}}^w$ can have non-vanishing value on the boundary $w=0$.

\section{Massless $p$-form Perturbations}
\label{sec:massless p-form}
In this section, we study massless abelian $p$-form gauge field perturbations on top of the background discussed in section~\ref{sec:background}.

As in the massless vector case discussed in section~\ref{sec:massless-vector}, we induce only $d$-dimensional perturbations and we impose $B^{\r M_2 \cdots M_p} =0$.
Therefore, the perturbation action is given by
	\begin{align}
		I_p \, = \, - \frac{1}{2(p+1)!} \int_M d^{d+1}x \sqrt{g} \, H^{\m_1 \cdots \m_{p+1}} H_{\m_1 \cdots \m_{p+1}} \, ,
	\label{I_p}
	\end{align}
with the field strength
	\begin{align}
		H_{M_1 \cdots M_{p+1}} \, = \, \frac{1}{p!} \, \nabla_{[M_1} B_{M_2 \cdots M_{p+1}]} \, ,
	\end{align}
where $[ \cdots ]$ is a totally anti-symmetrization without any numerical coefficients. 
The equation of motion derived from this action reads 
	\begin{align}
		\nabla_{M_1} H^{M_1 \cdots M_{p+1}} \, = \, 0 \, .
	\end{align}
Now we also impose the $d$-dimensional transverse gauge 
	\begin{align}
		\tilde{\nabla}_{\m_1} B^{\m_1 \cdots \m_p} \, = \, 0 \, .
	\label{transverse gauge}	
	\end{align}
We note that with the condition $B^{\r M_2 \cdots M_p} =0$, this $d$-dimensional transverse gauge also implies ($d+1$)-dimensional transverse gauge
	\begin{align}
		\nabla_{M_1} B^{M_1 \cdots M_p} \, = \, 0 \, .
	\end{align}
Upon this transverse gauge, the equation of motion is rewritten as
	\begin{align}
		\Big[\Box + \, p \, (d+1-p) \Big] B^{M_1 \cdots M_p} \, = \, 0 \, .
	\end{align}
Now we decompose the $\r$-direction. Then the equation of motion leads to
	\begin{align}
		\left[ a^{-2} (\tilde{\Box} + p ) + \pa_\r^2 + \, (d+2p) \tanh \r \, \pa_\r + \, 2pd - \, \frac{p(d+p-1)}{\cosh^2 \r} \right] B^{\m_1 \cdots \m_p} \, = \, 0 \, .
	\end{align}
We assume the separable form of the solutions
	\begin{align}
		B^{\m_1 \cdots \m_p}(\r, x, y) \, = \, a(\r)^{-p} \sum_{n=0}^\inf \, \mathcal{R}_n(\r) Y_n^{\m_1 \cdots \m_p}(x, y) \, .
	\end{align}
Then, the equation of motion is decomposed to the $\r$-direction and the remaining $d$-dimensional equations as 
	\begin{gather}
		\mathcal{R}_n'' \, + \, d \, \tanh \r \, \mathcal{R}_n' \, + \, p (d-p) \mathcal{R}_n \, = \, \frac{-\l_n}{\cosh^2 \r} \, \mathcal{R}_n \, , \label{eq_rho2} \\
		(\tilde{\Box} + p ) Y_n^{\m_1 \cdots \m_p} \, = \, \l_n Y_n^{\m_1 \cdots \m_p} \label{eq_x2} \, .
	\end{gather}

\subsection{$\r$-direction}
\label{sec:rho2}
The $\r$-direction equation (\ref{eq_rho2}) is solved by the same form as the massless vector case by the associated Legendre functions as
	\begin{align}
		\mathcal{R}_n(\r) \, = \, (1-\z^2)^{\frac{d}{4}} \left[ c_1 \, P_{\n-\frac{1}{2}}^\m(\z) \, + \, c_2 \, Q_{\n-\frac{1}{2}}^\m(\z) \right] \, ,
	\end{align}
where the order $\m$ is modified as
	\begin{align}
		\z \, = \, \tanh \r \, , \qquad
		\m \, = \, \frac{d-2p}{2} \, , \qquad
		\n \, = \, \frac{1}{2} \sqrt{(d-1)^2 + 4\l_n} \, .
	\label{pform-mu}
    \end{align}

For the asymptotic AdS boundary ($\z \to 1$ or $\r \to \inf$), we impose the standard Dirichlet boundary condition.
Then imposing this boundary condition, the solution is written as \cite{Izumi:2022opi}
	\begin{align}
		\mathcal{R}_n(\r) \, = \, B_0 \, (1-\z^2)^{\frac{d}{4}} \, P_{\n-\frac{1}{2}}^{-\m}(\z) \, ,
	\end{align}
where $B_0$ is a numerical coefficient.
This solution behaves as 
	\begin{align}
		\mathcal{R}_n(\r) \, \propto \, (1- \z)^{\frac{d-p}{2}} \, ,
	\end{align}
towards the asymptotic AdS boundary ($\z \to 1$).
Therefore, the behavior of the original gauge field is
	\begin{align}
		B^{\m_1 \cdots \m_p}(\r, x) \, \sim \, e^{-d \r} \, .
	\label{B-asympto}
	\end{align}

On the EOW brane, we impose Neumann boundary condition \cite{Takayanagi:2011zk, Fujita:2011fp}
	\begin{align}
		0 \, = \, H_{\r \m_1 \cdots \m_p} \big|_{\r = \r_*} \, .
	\end{align}
Since we set $B^{\r \m_2 \cdots \m_p}=0$, this condition implies 
	\begin{align}
		0 \, = \, \pa_\r \big( a^{2p} B^{\m_1 \cdots \m_p} \big) \big|_{\r = \r_*} \, ,
	\end{align}
or equivalently
	\begin{align}
		0 \, = \, \frac{d}{d\z} \big( a^p \, \mathcal{R}_n \big) \big|_{\z=\z_*} \, . 
	\end{align}
where $\z_*=\tanh \r_*$.
Using a property of the Legendre function, this condition can be written in a simply form as in the massless vector case:
	\begin{align}
		0 \, = \, P_{\n-\frac{1}{2}}^{1-\m}(\z_*) \, . 
	\label{BCs2}
	\end{align}
This condition quantizes the eigenvalue (or Kaluza-Klein mass) $\l_n$.

As in the massless vector case, we can find exact solutions when $d=2p+1$ and $d=2p+3$ cases.
For $d=2p+1$, the solutions are given by
	\begin{align}
		\n_n \, = \, \left( \frac{2n+1}{2} \right) \frac{\pi}{\th_*} \, , \qquad (n=0,1,2, \cdots )
	\end{align}
and for $d=2p+3$, the solutions are given by
	\begin{align}
		\n_n \, = \, \frac{(n+1)\pi}{\th_*} \, , \qquad (n=0,1,2, \cdots )
	\end{align}

\begin{figure}[t!]
	\begin{center}
		\scalebox{0.71}{\includegraphics{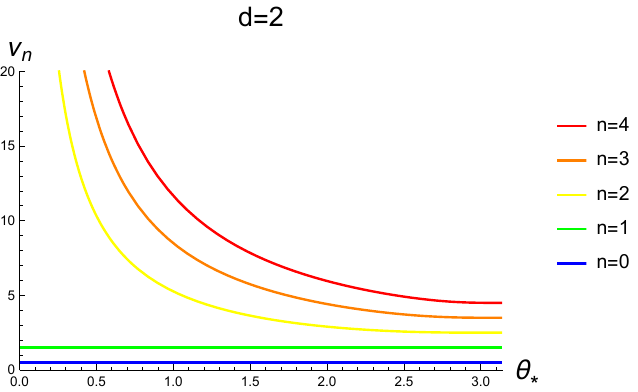}} \ \quad \scalebox{0.71}{\includegraphics{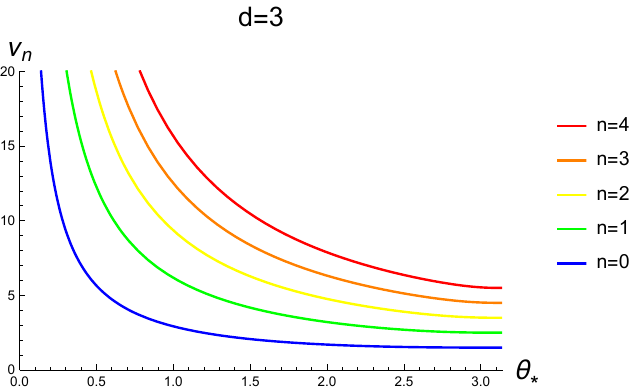}} \\[8pt]
		\scalebox{0.71}{\includegraphics{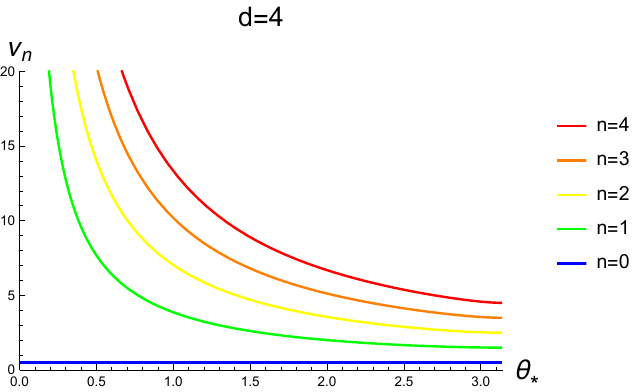}} \ \quad \scalebox{0.71}{\includegraphics{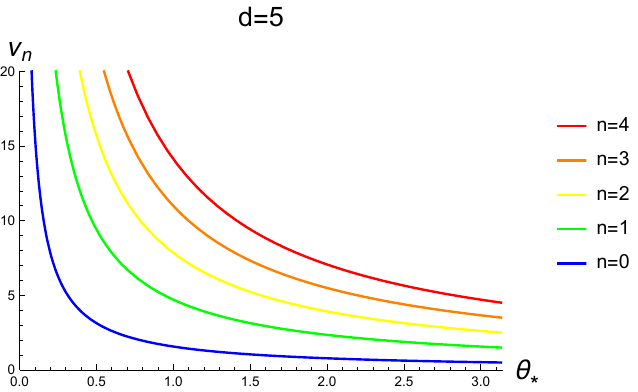}} 	
	\caption{Brane tension dependence of $p=2$ numerical solutions for the lowest five modes in AdS$_{d+1}$ with $d=2,3,4$ and $5$.}
	\label{fig:rho*_dependence(2)}
	\end{center}
\end{figure}

In figure~\ref{fig:rho*_dependence(2)}, for $p=2$ we show numerical solutions for low lying modes $\n_n$ with $n=0,1,2,3,4$.
We notice that, the lowest two mode ($n=0, 1$)  for $d=2$ and the lowest mode ($n=0$)  for $d=4$ are actually independent of brane location $\th_*$.
This can be again explained as follows, as in the vector case discussed in section~\ref{sec:rho}.
For $d=$ even case, we denote $d=2k$.
Then, the Neumann boundary condition (\ref{BCs2}) can be written as
	\begin{align}
		0 \, = \, P_{\n-\frac{1}{2}}^{1-k+p}(\z_*) \, = \, (-1)^{1+p-k} (1- \z_*^2)^{\frac{1+p-k}{2}} \left[ \frac{d^{1+p-k}}{d\z^{1+p-k}} \, P_{\n-\frac{1}{2}}(\z) \right] \bigg|_{\z=\z_*}\, . 
	\end{align}
Therefore, the first $1+p-k$ Legendre polynomials give the solutions.
Hence for $d=2k$, the lowest constant modes are given by
	\begin{align}
		\qquad \n_n \, = \, \frac{1}{2} + n  \, , \qquad ( n = 0, 1, 2, \cdots p-k, \ \ \textrm{for} \ \, d=2k)
	\end{align}

As far as this Neumann boundary condition on the brane is concerned, the gravitational perturbation studied in \cite{Izumi:2022opi} formally corresponds to the $p=0$ case.
Therefore, in the gravitational perturbation case, we do not find any brane-tension-independent mode.

The asymptotic spectrum for the present case is also given by (\ref{asymp_spec}), with $\m$ now given by (\ref{pform-mu}).

\subsection{$x$-direction}
\label{sec:x2}
Now we consider the remaining $d$-dimensional equation (\ref{eq_x2}).
We study this equation on the $d$-dimensional Poincare metric 
	\begin{align}
		ds_d^2 \, = \, \g_{\m\n} dq^\m dq^\n \, = \, \frac{dy^2 + \sum_{i=1}^{d-1} dx_i^2}{y^2} \, .
	\end{align}
We note that since $Y^{\m_1 \cdots \m_p}$ is totally antisymmetric, the non-zero components are $Y^{y i_2 \cdots i_p}$ and $Y^{i_1 \cdots i_p}$.
The transverse gauge (\ref{transverse gauge}) is explicitly written as
	\begin{align}
		\pa_i Y^{i \m_2 \cdots \m_p} \, = \, - \pa_y Y^{y\m_2 \cdots \m_p} \, + \, \frac{d}{y} \, Y^{y\m_2 \cdots \m_p} \, .
	\label{transverse gauge2}
	\end{align}
Using this condition, we can write the $x$ -direction equation (\ref{eq_x}) explicitly as
	\begin{align}
		0 \, &= \, \Box_0 Y^{yi_2 \cdots i_p} - \, \left( \frac{d+2p}{y} \right) \pa_y Y^{yi_2 \cdots i_p} + \left( \frac{(p+1)(d+p)-\l_n}{y^2} \right) Y^{yi_2 \cdots i_p} \, , \label{eq_Y^y2} \\
		0 \, &= \, \Box_0 Y^{i_1 \cdots i_p} - \, \left( \frac{d+2p-2}{y} \right) \pa_y Y^{i_1 \cdots i_p}
        - \frac{2}{y} \Big( \pa_{i_1} Y^{y i_2 \cdots i_p} + \cdots + \pa_{i_p} Y^{i_1 \cdots i_{p-1} y} \Big) \nn\\
		&\hspace{120pt}+ \, \left( \frac{p(d+p-1)-\l_n}{y^2} \right) Y^{i_1 \cdots i_p} \, , \label{eq_Y^i2} 
	\end{align}
where
	\begin{align}
		\Box_0 \, := \, \pa_y^2 + \pa_i \pa_i \, .
	\end{align}

The $Y^{yi_2 \cdots i_p}$ equation (\ref{eq_Y^y2}) is solved as
	\begin{align}
		Y^{yi_2 \cdots i_p}(x, y) \, = \, y^{\frac{d+2p+1}{2}} \left[ c_3 \, J_\n(\sqrt{-\bs{k}^2} y) \, + \, c_4 \, Y_\n(\sqrt{-\bs{k}^2} y) \right] e^{i \bs{k} \cdot \bs{x}} \, .
	\label{Y^y2}
	\end{align}
We impose Dirichlet boundary condition at $y=0$ as in \cite{Izumi:2022opi}, which sets $c_4=0$.
Therefore, the asymptotic behavior ($y \to 0$) is given by
	\begin{align}
		Y^{yi_2 \cdots i_p}(x, y) \, \sim \, y^{\frac{d+2p+1}{2}+\n} \, .
	\end{align}

For the remaining modes $Y^{i_1 \cdots i_p}$, as in the massless vector case, we decompose it into helicity-$(p-1)$ component $\phi^{i_1 \cdots i_{p-1}}$ and helicity-$p$ component $V^{i_1 \cdots i_p}$ as
	\begin{align}
		Y^{i_1 \cdots i_p} \, = \, \pa^{[i_1} \phi^{i_2 \cdots i_p]} \, + \, V^{i_1 \cdots i_p} \, ,
	\end{align}
where $V^{i_1 \cdots i_p}$ is totally antisymmetric and satisfies
	\begin{align}
		\pa_{i_1} V^{i_1 \cdots i_p} \, = \, 0 \, .
	\end{align}
By the transverse gauge (\ref{transverse gauge2}), the helicity-$(p-1)$ component $\phi^{i_1 \cdots i_{p-1}}$ must satisfy
	\begin{align}
		\pa_{i_1} \pa^{[i_1} \phi^{i_2 \cdots i_p]}\, = \, - \pa_y Y^{yi_2 \cdots i_p} \, + \, \frac{d}{y} \, Y^{yi_2 \cdots i_p} \, .
	\end{align}
Therefore, the asymptotic behavior of the helicity-$(p-1)$ component $Y_{(p-1)}^{i_1 \cdots i_p}=\pa^{[i_1} \phi^{i_2 \cdots i_p]}$ is given by
	\begin{align}
		Y_{(p-1)}^{i_1 \cdots i_p}(x, y) \, \sim \, y^{\frac{d+2p-1}{2}+\n} \, .
	\end{align}
For the helicity-$p$ component $Y_{(p)}^{i_1 \cdots i_p}=V^{i_1 \cdots i_p}$, $Y^{yi_2 \cdots i_p}$ does not contribute. Therefore, the equation (\ref{eq_Y^i2}) is simply solved as
	\begin{align}
		Y_{(p)}^{i_1 \cdots i_p}(x, y) \, = \, y^{\frac{d+2p-1}{2}} \left[ c_5 \, J_\n(\sqrt{-\bs{k}^2} y) \, + \, c_6 \, Y_\n(\sqrt{-\bs{k}^2} y) \right] e^{i \bs{k} \cdot \bs{x}} E^{i_1 \cdots i_p}(\bs{k}, s) \, ,
	\label{Y^i2}
	\end{align}
where $E^{i_1 \cdots i_p}(\bs{k}, s)$ is a basis tensor satisfying 
	\begin{align}
		k_{i_1} E^{i_1 \cdots i_p}(\bs{k}, s) \, = \, 0 \, , \qquad E_{i_1 \cdots i_p}(\bs{k}, s) E^{i_1 \cdots i_p}(\bs{k}, s') \, = \, \d_{ss'} \, .
	\end{align}
Here, $s$ is the label representing different choices of basis $p$-form.
Therefore, the asymptotic behavior of the helicity-$1$ component is same as that of helicity-$0$ component, so the transverse component has
	\begin{align}
		Y^{i_1 \cdots i_p}(x, y) \, \sim \, y^{\frac{d+2p-1}{2}+\n} \, .
	\end{align}

\subsection{Holographic Generalized Conserved Currents}
\label{sec:currents2}
The holographic conserved current $J^{\m_1 \cdots \m_p}$ is obtained from bulk gauge field perturbations by 
	\begin{align}
		J^{\m_1 \cdots \m_p}(x, w) \, = \, \lim_{z \to 0} \, z^{-d} B^{\m_1 \cdots \m_p}(x, w, z) \, ,
	\label{J^a(5)}
	\end{align}
in the Poincare coordinates. This formula is derived in Appendix~\ref{app:scaling}.
Using the coordinate transformations (\ref{coord-transf}), we can rewrite this in the hyperbolic slicing coordinates as
	\begin{align}
		J^{\m_1 \cdots \m_p}(x, w) \, = \, \lim_{\r \to \inf} \left[ \left( \frac{e^\r}{2y}\right)^d B^{\m_1 \cdots \m_p}(\r, x, y) \right] \Bigg|_{y=w}\, .
	\label{J^a(6)}
	\end{align}
Since the asymptotic behavior of the gauge field is given by (\ref{B-asympto})
	\begin{align}
		B^{\m_1 \cdots \m_p}(\r, x) \, \sim \, e^{-d \r} \, .
	\end{align}
the $\r \to \inf$ limit in (\ref{J^a(6)}) indeed gives a finite value.
Finally combining with the asymptotic behavior of $Y^{\m_1 \cdots \m_p}$ discussed in section~\ref{sec:x2}, near-the-boundary behavior of the holographic conserved currents are found as
	\begin{align}
		J^{wi_2 \cdots i_p}(x, w) \, \propto \, w^{\, \n+\frac{-d+2p+1}{2}} \, , \qquad J^{i_1 \cdots i_p}(x, w) \, \propto \, w^{\, \n+\frac{-d+2p-1}{2}} \, .
	\end{align}
Since $\l_n$ is non-negative, we have
	\begin{align}
		\n \, \ge \, \frac{d-1}{2} \, .
	\end{align}
This implies that at least the powers of the perpendicular direction $w$ are given by
	\begin{align}
		J^{wi_2 \cdots i_p}(x, w) \, \propto \, w^{\, p} \, , \qquad J^{i_1 \cdots i_p}(x, w) \, \propto \, w^{\, p-1} \, .
	\end{align}
Therefore, the perpendicular component of the holographic conserved current $J^{wi_2 \cdots i_p}$ vanishes at the boundary $J^{wi_2 \cdots i_p}(w=0)=0$.
For the transverse components, $p=1$ seems a special case, for which one can have finite non-vanishing values at the boundary $w=0$,
while for $p\ge2$, our result indicates that the transverse components also vanishes at the boundary $J^{i_1 \cdots i_p}(w=0)=0$.
It would be interesting to further investigate whether this behavior is an artifact of our setup or general feature of a $p$-form operator in BCFT.

\section{Massive $p$-form Perturbations}
\label{sec:massive p-form}
In this section, we study massive $p$-form field perturbations on top of the background discussed in section~\ref{sec:background}.
Massive $p$-form fields in empty AdS were previously discussed in \cite{Arutyunov:1998xt, lYi:1998trg}.

Unlike the massless case, we cannot impose the gauge condition $B^{\r M_2 \cdots M_p}=0$, since there is no gauge symmetry in the this case.
Therefore, we induce the full $(d+1)$-dimensional $p$-form perturbations and the perturbation action is given by
	\begin{align}
		I_{\textrm{mp}} \, = \, - \frac{1}{2p!} \int_M d^{d+1}x \sqrt{g} \left[ \, \frac{1}{(p+1)} \, H^{M_1 \cdots M_{p+1}} H_{M_1 \cdots M_{p+1}} + m^2 \, B^{M_1 \cdots M_p} B_{M_1 \cdots M_p} \right] \, .
	\end{align}
The equation of motion derived from this action is 
	\begin{align}
		\nabla_N H^{N M_1 \cdots M_p} \, = \, m^2 B^{M_1 \cdots M_p} \, .
	\end{align}
Even though we do not have any gauge fixing condition, due to the anti-symmetric nature of $H^{M_1 \cdots M_{p+1}}$, we get a supplemental condition \cite{Arutyunov:1998xt, lYi:1998trg}
	\begin{align}
		\nabla_{M_1} B^{M_1 \cdots M_p} \, = \, 0 \, ,
	\label{sup-cond2}
	\end{align}
from the equation of motion.
Using this supplemental condition, the equation of motion is rewritten as
	\begin{align}
		\Big[\Box + \, p \, (d+1-p) - m^2 \Big] B^{M_1 \cdots M_p} \, = \, 0 \, .
	\label{massive-eq2}
	\end{align}

As in the massless case, once we separate the $\r$-direction, the equations of motion read
	\begin{align}
		\left[ a^{-2} \, \big( \tilde{\Box} + p -1 \big) + \, \pa_\r^2 + \big(d+2p\big) \tanh \r \, \pa_\r + 2pd - m^2 - \frac{p(d+p-1)}{\cosh^2 \r} \right] B^{\r\m_2 \cdots \m_p} \, = \, 0 \, , 
	\label{eq-B^rho}
	\end{align}
and 
	\begin{align}
		&\left[ a^{-2} \big(\tilde{\Box} + p \big) + \pa_\r^2 + \, \big(d+2p\big) \tanh \r \, \pa_\r + 2pd - m^2 - \frac{p(d+p-1)}{\cosh^2 \r} \right] B^{\m_1 \cdots \m_p} \nn\\
		&\hspace{100pt}  = \, -2 \tanh \r \Big( \pa^{\m_1} B^{\r \m_2 \cdots \m_p} + \pa^{\m_2} B^{\m_1 \r \m_3 \cdots \m_p} + \cdots \Big) \, .
	\label{eq-B^mu}
	\end{align}
To write these equations, we also used the supplemental condition (\ref{sup-cond2}), whose $\r$-separated form is given by
	\begin{align}
		\tilde{\nabla}_\m B^{\m M_2 \cdots M_p} \, = \, -\pa_\r B^{\r M_2 \cdots M_p} \, - \, d \tanh \r \, B^{\r M_2 \cdots M_p} \, .
	\end{align}
As in the massive vector case, we have a non-trivial equation for $B^{\r\m_2 \cdots \m_p}$.
Then the second equation is an inhomogeneous differential equation, which is sourced by $B^{\r \m_2 \cdots \m_p}$ in the right-hand side.

\subsection{Solution for $B^{\r\m_2 \cdots \m_p}$}
\label{sec:B^rho}
We first solve the equation (\ref{eq-B^rho}) for $B^{\r\m_2 \cdots \m_p}$.
As in the massless case, we assume the separable form of the solution and take an ansatz
	\begin{align}
		B^{\r\m_2 \cdots \m_p}(\r, x, y) \, = \, a(\r)^{-p} \sum_{n=0}^\inf \, \mathcal{R}_n(\r) Y^{\r\m_2 \cdots \m_p}_n(x, y) \, .
	\label{B^rho}
	\end{align}
Then, the equation of motion is decomposed as 
	\begin{gather}
		\mathcal{R}_n'' \, + \, d \, \tanh \r \, \mathcal{R}_n' \, + \, \big(p(d-p)-m^2\big) \mathcal{R}_n \, = \, \frac{-\l_n}{\cosh^2 \r} \, \mathcal{R}_n \, , \\
		\big( \tilde{\Box} + p - 1 \big) \,Y_n^{\r\m_2 \cdots \m_p} \, = \, \l_n Y_n^{\r\m_2 \cdots \m_p} \, .
	\end{gather}

The equation for $\mathcal{R}_n$ is basically the same structure as in the massless case, so the solution is given by
	\begin{align}
		\mathcal{R}_n(\r) \, = \, B_0 \, (1-\z^2)^{\frac{d}{4}} \, P_{\n-\frac{1}{2}}^{-\m}(\z) \, ,
	\label{R_n-solution2}
	\end{align}
but the order of the associated Legendre function is now modified by the mass as
	\begin{align}
		\z \, = \, \tanh \r \, , \qquad
		\m \, = \, \frac{1}{2} \sqrt{(d-2p)^2 + 4m^2} \, , \qquad
		\n \, = \, \frac{1}{2} \sqrt{(d-1)^2 + 4\l_n} \, .
	\label{massive-mu2}
	\end{align}
This solution behaves as 
	\begin{align}
		\mathcal{R}_n(\r) \, \sim \, (1- \z)^{\frac{d+2\m}{4}} \, ,
	\end{align}
towards the asymptotic AdS boundary ($\z \to 1$).
Therefore, the behavior of the original gauge field is
	\begin{align}
		B^{\r\m_2 \cdots \m_p} \, \sim \, e^{- (\frac{d+2p}{2} + \m ) \r} \, .
	\label{B-asympto2}
	\end{align}
Once we impose a boundary condition on the EOW brane at $\r = \r_*$, this condition quantizes the eigenvalue $\l_n$.

Next for the $Y_n^{\r\m_2 \cdots \m_p}$, we need to consider $Y_n^{\r y i_3 \cdots i_p}$ and $Y_n^{\r i_2 \cdots i_p}$.
Their equations of motion are
	\begin{align}
		0 \, &= \, \Box_0 Y^{\r y i_3 \cdots i_p} - \, \left( \frac{d+2p-2}{y} \right) \pa_y Y^{\r y i_3 \cdots i_p} + \, \left( \frac{p(d+p-1) - \l_n}{y^2} \right) \, Y^{\r y i_3 \cdots i_p} \, , \\
		0 \, &= \, \Box_0 Y^{\r i_2 \cdots i_p} - \, \left( \frac{d+2p-4}{y} \right) \pa_y Y^{\r i_2 \cdots i_p} - \, \frac{2}{y} \Big( \pa^{i_2} Y^{\r y i_3 \cdots i_p} + \pa^{i_3} Y^{\r i_2 y i_3 \cdots i_p} + \cdots \Big) \nn\\
		&\qquad \qquad + \, \left( \frac{(p-1)(d+p-2) - \l_n}{y^2} \right) \, Y^{\r i_2 \cdots i_p} \, .
	\end{align}
These equations are identical to (\ref{eq_Y^y2}) and (\ref{eq_Y^i2}) with replacement of $p \to p-1$.
Therefore, the solutions are also given in section~\ref{sec:x2} with replacement of $p \to p-1$.
In particular their asymptotic behaviors ($y \to 0$) are given by
	\begin{align}
		Y^{\r y i_3 \cdots i_p}(x, y) \, \sim \, y^{\frac{d+2p-1}{2}+\n} \, , \qquad Y^{\r i_2 \cdots i_p}(x, y) \, \sim \, y^{\frac{d+2p-3}{2}+\n} \, .
	\label{Y^rho2}
	\end{align}

\subsection{Solution for $B^{\m_1 \cdots \m_p}$}
\label{sec:B^rho}
For the equation (\ref{eq-B^mu}) for $B^{\m_1 \cdots \m_p}$, we first study the homogeneous equation.
For the homogeneous solution, we assume the separable form of the solution and take an ansatz
	\begin{align}
		B_{\textrm{hom}}^{\m_1 \cdots \m_p}(\r, x, y) \, = \, a(\r)^{-1} \sum_{n=0}^\inf \, \mathcal{R}_n(\r) Y_n^{\m_1 \cdots \m_p}(x, y) \, .
	\end{align}
Then, the equation of motion is decomposed as 
	\begin{gather}
		\mathcal{R}_n'' \, + \, d \, \tanh \r \, \mathcal{R}_n' \, + \, \big(p(d-p)-m^2\big) \mathcal{R}_n \, = \, \frac{-\l_n}{\cosh^2 \r} \, \mathcal{R}_n \, , \\
		(\tilde{\Box} + p ) \, Y_n^{\m_1 \cdots \m_p} \, = \, \l_n Y_n^{\m_1 \cdots \m_p} \, . 
	\end{gather}
We can see that the equation for $\mathcal{R}_n$ is the same as $B^{\r\m_2 \cdots \m_p}$ case discussed in the previous subsection. Therefore, the solution is given by (\ref{R_n-solution2}) with (\ref{massive-mu2}).
On the other hand, the equation for $Y_n^{\m_1 \cdots \m_p}$ is the same as massless case discussed in the previous section, provided that the eigenvalue $\l_n$ is appropriately quantized in the current massive case. 
Therefore, besides the precise quantization condition for $\l_n$, the solutions are given by (\ref{Y^y2}) and (\ref{Y^i2}) with $c_4=c_6=0$.
In particular, the asymptotic behavior ($y \to 0$) of the homogeneous solutions are given by
	\begin{align}
		Y_{\textrm{hom}}^{y i_2 \cdots i_p}(x, y) \, \sim \, y^{\frac{d+2p+1}{2}+\n} \, , \qquad Y_{\textrm{hom}}^{i_1 \cdots i_p}(x, y) \, \sim \, y^{\frac{d+2p-1}{2}+\n} \, .
	\label{homo-sol2}
	\end{align}

Next, we consider an inhomogeneous solution for (\ref{eq-B^mu}).
It looks difficult to find an exact solution, since the solution of $A^\r$ is given in the summation form as in (\ref{B^rho}).
Therefore, in this subsection, let us just determine the leading asymptotic behavior of an inhomogeneous solution.
For the $\r$ direction, we use the fooling ansatz for the special solution
	\begin{align}
		\hat{B}^{\m_1 \cdots \m_p}(\r, x, y) \, = \, e^{- (\frac{d+2p}{2} + \m ) \r} \, \hat{Y}^{\m_1 \cdots \m_p}(x, y) \, .
	\end{align}
We use a hat to denote this special solution. With this ansatz, the leading contribution (in large $\r$) of the inhomogeneous equation (\ref{eq-B^mu}) identically vanishes and the subleading contribution is given by
	\begin{align}
		\big( \tilde{\Box} + p - \hat{\l} \big) \hat{Y}^{\m_1 \cdots \m_p} \, = \, -2 \big( \pa_{\m_1} Y^{\r \m_2 \cdots \m_p} + \cdots \big) \, ,
	\end{align}
where we defined 
	\begin{align}
		\hat{\l} \, := \, p(d+p-1) - \left( \frac{d+2p}{2} \right) \left( \frac{d+2p}{2} + \m \right) \, .
	\end{align}
With this definition of $\hat{\l}$, the LHS becomes the same form as (\ref{eq_x2}), whose explicit form is given by (\ref{eq_Y^y2}) and (\ref{eq_Y^i2}).
Therefore for $\hat{Y}^{\m_1 \cdots \m_p}$, we need to solve 
	\begin{align}
		&\Box_0 \hat{Y}^{yi_2 \cdots i_p} - \, \left( \frac{d+2p}{y} \right) \pa_y \hat{Y}^{yi_2 \cdots i_p} + \, \left( \frac{(p+1)(d+p)-\hat{\l}}{y^2} \right) \hat{Y}^{yi_2 \cdots i_p} \\
		&\qquad = \, - \frac{2}{y^2} \Big( \pa_y Y^{\r i_2 \cdots i_p} + \cdots \Big)\, , \nn\\
		&\Box_0 \hat{Y}^{i_1 \cdots i_p} - \, \left( \frac{d+2p-2}{y} \right) \, \pa_y \hat{Y}^{i_1 \cdots i_p} + \, \left( \frac{p(d+p-1)-\hat{\l}}{y^2} \right) \hat{Y}^{i_1 \cdots i_p} \\
		&\qquad  = \, \frac{2}{y} \Big( \pa_{i_1} \hat{Y}^{yi_2 \cdots i_p} + \cdots \Big) - \, \frac{2}{y^2} \Big( \pa_{i_1} Y^{\r i_2 \cdots i_p} + \cdots \Big) \, . \nn
	\end{align}
Comparing with the asymptotic behavior of $Y^{\r \m_2 \cdots \m_p}$ (\ref{Y^rho2}), we find the asymptotic behavior of the special solution as
	\begin{align}
		\hat{Y}^{yi_2 \cdots i_p} \, \sim \, y^{\frac{d+2p-5}{2}+\n} \, , \qquad \hat{Y}^{i_1 \cdots i_p} \, \sim \, y^{\frac{d+2p-3}{2}+\n}  \, .
	\end{align}
Therefore for all components of $\hat{Y}^\m$, we have found that the exponents of the asymptotic behavior are smaller than those of the general solution for the homogeneous equation (\ref{homo-sol2}).

\subsection{Dual $p$-form Operators}
The dual $p$-form operator $\mathcal{O}^{\m_1 \cdots \m_p}$ is obtained from bulk massive $p$-form field perturbation by 
	\begin{align}
		\mathcal{O}^{\m_1 \cdots \m_p}(x, w) \, = \, \lim_{z \to 0} \, z^{-\left( \frac{d+2p}{2} + \m \right)} B^{\m_1 \cdots \m_p}(x, w, z) \, ,
	\label{J^a(7)}
	\end{align}
in the Poincare coordinates. This formula is derived in Appendix~\ref{app:scaling}.
Using the coordinate transformations (\ref{coord-transf}), we can rewrite this in the hyperbolic slicing coordinates as
	\begin{align}
		\mathcal{O}^{\m_1 \cdots \m_p}(x, w) \, = \, \lim_{\r \to \inf} \left[ \left( \frac{e^\r}{2y}\right)^{\frac{d+2p}{2} + \m} B^{\m_1 \cdots \m_p}(\r, x, y) \right] \Bigg|_{y=w}\, .
	\label{J^a(8)}
	\end{align}
Since the asymptotic behavior of the massive $p$-form field is given by (\ref{A-asympto2})
	\begin{align}
		B^{\m_1 \cdots \m_p}(\r, x, y) \, \sim \, e^{- (\frac{d+2p}{2} + \m ) \r} \, ,
	\end{align}
for both the homogeneous and inhomogeneous solutions, the $\r \to \inf$ limit in (\ref{J^a(7)}) indeed gives a finite value.
Therefore, we next need to consider the asymptotic behavior of $Y^{\m_1 \cdots \m_p}$.
Let us separately consider the contributions from the homogeneous and inhomogeneous solutions.
First, the contribution from the homogeneous solution is found as
	\begin{align}
		\mathcal{O}_{\textrm{hom}}^{wi_2 \cdots i_p}(x, w) \, \propto \, w^{\, \n+\frac{1}{2}-\m} \, , \qquad \mathcal{O}_{\textrm{hom}}^{i_1 \cdots i_p}(x, w) \, \propto \, w^{\, \n-\frac{1}{2}-\m} \, .
	\end{align}
If we take the massless limit $m^2 \to 0$, these scaling behaviors coincide with those of the conserved currents, discussed in section \ref{sec:currents2}.
Therefore, in the massless limit, $\mathcal{O}_{\textrm{hom}}^{wi_2 \cdots i_p}$ vanishes on the boundary $w=0$.

Next we show that the contribution from the inhomogeneous solution gives qualitatively different behavior compared to these contributions from the homogeneous solutions.
The contribution from the inhomogeneous solution is found as
	\begin{align}
		\mathcal{O}_{\textrm{inh}}^{wi_2 \cdots i_p}(x, w) \, \propto \, w^{\, \n-\frac{5}{2}-\m} \, , \qquad \mathcal{O}_{\textrm{inh}}^{i_1 \cdots i_p}(x, w) \, \propto \, w^{\, \n-\frac{3}{2}-\m} \, .
	\end{align}
Even in the massless limit, $\mathcal{O}_{\textrm{inh}}^{wi_2 \cdots i_p}$ can have a negative power in $w$ in the close to the boundary limit.
Therefore, in general the contribution from the inhomogeneous solution $\mathcal{O}_{\textrm{inh}}^{wi_2 \cdots i_p}$ can have non-vanishing value on the boundary $w=0$.

\section{Conclusions and discussions}
\label{sec:conclusions}
In this paper, we studied massless/massive vector and $p$-form field perturbations in the bulk of the AdS/BCFT setup.
We employed $U(1)$ preserving Neumann boundary condition on the end-of-the-world brane. 
For massless perturbations, we studied sepectrums carefully, and found several brane-tension-independent modes,
which are understood as boundary-condition-independent modes in the dual BCFT.
For massless cases, by imposing Lorentz gauge, the equations of motion can be decomposed into each component, and they become homogeneous differential equations.
The general solution of the perpendicular component leads to the vanishing-on-the-boundary behavior (\ref{J_int}) of the holographic conserved currents.
On the other hand, for massive perturbations, the equations of motion {\it cannot} be decomposed into each component, and they are {\it inhomogeneous} differential equations.
This is also true in the massless limit, due to the absence of gauge symmetry.
The general solutions of the corresponding homogeneous equation are simple massive generalization of the massless solutions,
whose perpendicular component leads to the vanishing-on-the-boundary behavior (\ref{J_int}) in the massless limit.
However, a special solution of the inhomogeneous equation leads to the non-vanishing-on-the-boundary behavior (\ref{O_int}) even in the massless limit.
Most part of this paper is rather technical, but we hope that some of the results would be useful for further understanding of the AdS/BCFT correspondence.
Some possible future directions are as follows.

In this paper, we studied a $U(1)$ preserving Neumann boundary condition, but it is also interesting to consider $U(1)$ breaking Neumann boundary condition \cite{Gaberdiel:2001zq}.
For this purpose, we probably need a coupling to brane-localized fields or currents, as in the scalar field cases studied in \cite{Suzuki:2022xwv, Izumi:2022opi, Kanda:2023zse}.
Also from the perspective of causality \cite{Omiya:2021olc, Mori:2023swn}, such brane-localized fields are perhaps necessary in the AdS/BCFT setup.

We found several brane-tension-independent modes in massless perturbations in $d=$ even dimensions.
Since if we turn on masses slightly, these modes immediately disappear, it is tempting to think that these modes are somehow related to the global symmetry of the dual BCFTs.
It would be particularly nice if we could identify these modes from the boundary conformal bootstrap \cite{McAvity:1995zd, Liendo:2012hy, Kaviraj:2018tfd, Mazac:2018biw, Kusuki:2021gpt, Kusuki:2022ozk}.

In general, $p$-form conserved currents (like what we studied in section~\ref{sec:currents2}) generate higher-form symmetries \cite{Gaiotto:2014kfa,Gomes:2023ahz,Bhardwaj:2023kri}.
Some type of higher-form symmetries have been discussed in the context of the AdS/CFT correspondence (e.g. \cite{Bergman:2020ifi,Hofman:2017vwr}).
However, up to the author's knowledge, higher-form symmetries have not yet discussed in the context of the AdS/BCFT correspondence.
It would be interesting to clarify meanings of higher-form symmetries in AdS/BCFT ant its relation to the EOW brane.
We believe that our studies on the $p$-form gauge field perturbation will be also useful for this purpose.

\section*{Acknowledgements}

We are grateful to Tadashi Takayanagi and Tomonori Ugajin for useful comments on the draft of this paper.
This work is supported by JSPS KAKENHI Grant No.~23K13105.

\appendix
\section{Scaling Behaviors}
\label{app:scaling}

\subsection{Vector Case}
\label{sec:vector}
In this appendix, we consider the asymptotic scaling behavior of the perturbations.
Since for this purpose, massless case is simply obtained by taking $m^2 \to 0$ limit from the massive case, we consider the massive vector in this appendix.
It is enough to study the equation of motion (\ref{massive-eq})
	\begin{align}
		0 \, = \, (\Box + d - m^2 ) A^M \, ,
	\end{align}
in the Poincare coordinates (\ref{Poincare})
	\begin{align}
		ds_{d+1}^2 \, = \, \frac{dz^2+dw^2 + \sum_{i=1}^{d-2} dx_i^2}{z^2} \, .
	\end{align}

The $M=z$ component of the equation of motion is explicitly written as
	\begin{align}
		0 \, = \, z^2 \pa_N \pa_N A^z \, - \, (d+3) z \, \pa_z A^z \, + \, (3d+3-m^2) A^z \, .
	\end{align}
Here we are only concerned of the asymptotic behavior ($z \to 0$), substituting a scaling ansazt 
	\begin{align}
		A^z \, \sim \, z^\a \, ,
	\end{align}
the equation of motion gives 
	\begin{align}
		0 \, = \, \a (\a - d - 4) + d + 3 -m^2 \, ,
	\end{align}
and its solutions are given by
	\begin{align}
		\a_\pm \, = \, \frac{d+4}{2} \, \pm \, \frac{1}{2} \sqrt{(d-2)^2 + 4m^2} \, .
	\end{align}
Hence, the asymptotic behavior of $A^z$ is found as
	\begin{align}
		A^z \, \sim \, A^z_{(-)} z^{\a_-} + \, J^z \, z^{\a_+} \, .
	\end{align}

Next, the $M=\m$ component of the equation of motion is explicitly written as
	\begin{align}
		z^2 \pa_N \pa_N A^\m \, - \, (d+1) z \, \pa_z A^\m \, + \, (2d-m^2) A^\m \, = \, 2z^{-1} \pa^\m A^z \, .
	\label{A_z-eq}
	\end{align}
This equation is an inhomogeneous differential equation and it has a source contribution from $A^z$ in the right-hand side.
Let us first consider the homogeneous solution.
Substituting a scaling ansazt 
	\begin{align}
		A^\m \, \sim \, z^\b \, ,
	\end{align}
into the homogeneous equation gives 
	\begin{align}
		0 \, = \, \b (\b - d - 2) + 2d -m^2 \, .
	\end{align}
The solutions are given by
	\begin{align}
		\b_\pm \, = \, \frac{d+2}{2} \, \pm \, \frac{1}{2} \sqrt{(d-2)^2 + 4m^2} \, .
	\end{align}
Now we consider an inhomogeneous solution.
Again just considering the asymptotic behavior, the inhomogeneous solution must match with the scaling of the RHS of (\ref{A_z-eq}).
This comparison gives the scaling behavior of the inhomogeneous solution as $z^{\a+1}$. This behavior is subleading compared with the scaling behavior of the inhomogeneous solution $z^\b$.
Therefore, the leading asymptotic behavior of $A^\m$ is determined by $\b_{\pm}$ as
	\begin{align}
		A^\m \, \sim \, A^\m_{(-)} z^{\b_-} + \, J^\m \, z^{\b_+} \, .
	\label{A^mu}
	\end{align}
Hence, the dual conserved currents are found by 
	\begin{align}
		J^\m \, = \, \lim_{z\to 0} z^{- \b_+} A^\m \, .
	\label{J^mu}
	\end{align}
This gives the formulae (\ref{J^a(1)}) and (\ref{J^a(3)}).

To be precise, for the $J^y$ component, we also need to consider the contribution from $A^z$.
Under the coordinate transformation (\ref{coord-transf}), the vector fields are related by 
	\begin{align}
		A^y \, = \, \frac{\pa y}{\pa w} \, A^w \, + \, \frac{\pa y}{\pa z} \, A^z \, .
	\end{align}
Since $y = \sqrt{w^2+z^2}$, this relation is explicitly given by
	\begin{align}
		A^y \, = \, \frac{1}{\sqrt{w^2+z^2}} \Big( w A^w + z A^z \Big) \, .
	\end{align}
Therefore, even for $A^y$, the asymptotic scaling behavior is same as those of $A^w$, which is included in (\ref{A^mu}).
Thus the $J^y$ component is also obtaind by the formula (\ref{J^mu}).

\subsection{$p$-form Case}
\label{sec:p-form}
In this appendix, we consider the asymptotic scaling behavior of the perturbations.
Since for this purpose, massless case is simply obtained by taking $m^2 \to 0$ limit from the massive case, we consider the massive $p$-form field in this appendix. (The vector case is simply obtained by setting $p=1$.)
It is enough to study the equation of motion (\ref{massive-eq2})
	\begin{align}
		\Big[\Box + \, p \, (d+1-p) - m^2 \Big] B^{M_1 \cdots M_p} \, = \, 0 \, ,
	\end{align}
in the Poincare coordinates (\ref{Poincare})
	\begin{align}
		ds_{d+1}^2 \, = \, \frac{dz^2+dw^2 + \sum_{i=1}^{d-2} dx_i^2}{z^2} \, .
	\end{align}

The $\{z, \m_2 , \cdots , \m_p\}$ component of the equation of motion is explicitly written as
	\begin{align}
		0 \, = \, \Big[ z^2 \pa_N \pa_N  \, - \, (d+2p+1) z \, \pa_z \, + \, (2p+1)(d+1) \, - \, m^2 \Big] B^{z \m_2 \cdots \m_p} \, .
	\end{align}
Here we are only concerned of the asymptotic behavior ($z \to 0$), so substituting a scaling ansazt 
	\begin{align}
		B^{z \m_2 \cdots \m_p}  \, \sim \, z^\a \, ,
	\end{align}
the equation of motion gives 
	\begin{align}
		0 \, = \, \a (\a - d - 2p - 2) + (2p+1)(d + 1) - m^2 \, ,
	\end{align}
and its solutions are given by
	\begin{align}
		\a_\pm \, = \, \frac{d+2p+2}{2} \, \pm \, \frac{1}{2} \sqrt{(d-2p)^2 + 4m^2} \, .
	\end{align}
Hence, the asymptotic behavior of $B^{z \m_2 \cdots \m_p}$ is found as
	\begin{align}
		B^{z \m_2 \cdots \m_p} \, \sim \, B^{z \m_2 \cdots \m_p}_{(-)} z^{\a_-} + \, J^{z \m_2 \cdots \m_p} \, z^{\a_+} \, .
	\end{align}

Next, the $\{\m_1 , \cdots , \m_p\}$ component of the equation of motion is explicitly written as
	\begin{align}
		&\Big[ z^2 \pa_N \pa_N \, - \, (d+2p-1) z \, \pa_z \, + \, 2pd-m^2 \Big] B^{\m_1 \cdots \m_p} \nn\\
        &\hspace{120pt} = \, 2z^{-1} \Big( \pa^{\m_1} B^{z \m_2 \cdots \m_p} + \cdots + \pa^{\m_p} B^{\m_1 \cdots \m_{p-1} z} \Big) \, .
	\label{B_z-eq}
	\end{align}
This equation is an inhomogeneous differential equation and it has a source contribution from $B^{z \m_2 \cdots \m_p}$ in the right-hand side.
Let us first consider the homogeneous solution.
Substituting a scaling ansazt 
	\begin{align}
		B^{\m_1 \cdots \m_p} \, \sim \, z^\b \, ,
	\end{align}
into the homogeneous equation gives 
	\begin{align}
		0 \, = \, \b (\b - d - 2p) + 2pd -m^2 \, .
	\end{align}
The solutions are given by
	\begin{align}
		\b_\pm \, = \, \frac{d+2p}{2} \, \pm \, \frac{1}{2} \sqrt{(d-2p)^2 + 4m^2} \, .
	\end{align}
Now we consider an inhomogeneous solution.
Again just considering the asymptotic behavior, the inhomogeneous solution must match with the scaling of the RHS of (\ref{B_z-eq}).
This comparison gives the scaling behavior of the inhomogeneous solution as $z^{\a+1}$. This behavior is subleading compared with the scaling behavior of the inhomogeneous solution $z^\b$.
Therefore, the leading asymptotic behavior of $B^{z \m_2 \cdots \m_p}$ is determined by $\b_{\pm}$ as
	\begin{align}
		B^{\m_1 \cdots \m_p} \, \sim \, B^{\m_1 \cdots \m_p}_{(-)} z^{\b_-} + \, J^{\m_1 \cdots \m_p} \, z^{\b_+} \, .
	\label{B^mu}
	\end{align}
Hence, the dual conserved currents are found by 
	\begin{align}
		J^{\m_1 \cdots \m_p} \, = \, \lim_{z\to 0} z^{- \b_+} B^{\m_1 \cdots \m_p} \, .
	\label{J^mu2}
	\end{align}
This gives the formulae (\ref{J^a(5)}) and (\ref{J^a(7)}).

To be precise, for the $J^{y\m_2 \cdots \m_p}$ components, we also need to consider the contribution from $B^{z \m_2 \cdots \m_p}$.
Under the coordinate transformation (\ref{coord-transf}), the vector fields are related by 
	\begin{align}
		B^{y \m_2 \cdots \m_p} \, = \, \frac{\pa y}{\pa w} \, B^{w \m_2 \cdots \m_p} \, + \, \frac{\pa y}{\pa z} \, B^{z \m_2 \cdots \m_p} \, .
	\end{align}
Since $y = \sqrt{w^2+z^2}$, this relation is explicitly given by
	\begin{align}
		B^{y \m_2 \cdots \m_p} \, = \, \frac{1}{\sqrt{w^2+z^2}} \Big( w B^{w \m_2 \cdots \m_p} + z B^{z \m_2 \cdots \m_p} \Big) \, .
	\end{align}
Therefore, even for $B^{y \m_2 \cdots \m_p}$, the asymptotic scaling behavior is same as those of $B^{w \m_2 \cdots \m_p}$, since the contribution from $B^{z \m_2 \cdots \m_p}$ is subleading.
This is included in (\ref{B^mu}), so the $J^{y \m_2 \cdots \m_p}$ component is also obtaind by the formula (\ref{J^mu2}).

\bibliographystyle{JHEP}
\bibliography{Refs}

\providecommand{\href}[2]{#2}\begingroup\raggedright\begin{thebibliography}{10}

\bibitem{Cardy:1984bb}
J.L.~Cardy, \emph{{Conformal Invariance and Surface Critical Behavior}},
  \href{https://doi.org/10.1016/0550-3213(84)90241-4}{\emph{Nucl. Phys. B}
  {\bfseries 240} (1984) 514}.

\bibitem{Cardy:2004hm}
J.L.~Cardy, \emph{{Boundary conformal field theory}},
  \href{https://arxiv.org/abs/hep-th/0411189}{{\ttfamily hep-th/0411189}}.

\bibitem{McAvity:1993ue}
D.M.~McAvity and H.~Osborn, \emph{{Energy momentum tensor in conformal field
  theories near a boundary}},
  \href{https://doi.org/10.1016/0550-3213(93)90005-A}{\emph{Nucl. Phys. B}
  {\bfseries 406} (1993) 655}
  [\href{https://arxiv.org/abs/hep-th/9302068}{{\ttfamily hep-th/9302068}}].

\bibitem{McAvity:1995zd}
D.M.~McAvity and H.~Osborn, \emph{{Conformal field theories near a boundary in
  general dimensions}},
  \href{https://doi.org/10.1016/0550-3213(95)00476-9}{\emph{Nucl. Phys. B}
  {\bfseries 455} (1995) 522}
  [\href{https://arxiv.org/abs/cond-mat/9505127}{{\ttfamily
  cond-mat/9505127}}].

\bibitem{Takayanagi:2011zk}
T.~Takayanagi, \emph{{Holographic Dual of BCFT}},
  \href{https://doi.org/10.1103/PhysRevLett.107.101602}{\emph{Phys. Rev. Lett.}
  {\bfseries 107} (2011) 101602}
  [\href{https://arxiv.org/abs/1105.5165}{{\ttfamily 1105.5165}}].

\bibitem{Fujita:2011fp}
M.~Fujita, T.~Takayanagi and E.~Tonni, \emph{{Aspects of AdS/BCFT}},
  \href{https://doi.org/10.1007/JHEP11(2011)043}{\emph{JHEP} {\bfseries 11}
  (2011) 043} [\href{https://arxiv.org/abs/1108.5152}{{\ttfamily 1108.5152}}].

\bibitem{Randall:1999ee}
L.~Randall and R.~Sundrum, \emph{{A Large mass hierarchy from a small extra
  dimension}}, \href{https://doi.org/10.1103/PhysRevLett.83.3370}{\emph{Phys.
  Rev. Lett.} {\bfseries 83} (1999) 3370}
  [\href{https://arxiv.org/abs/hep-ph/9905221}{{\ttfamily hep-ph/9905221}}].

\bibitem{Randall:1999vf}
L.~Randall and R.~Sundrum, \emph{{An Alternative to compactification}},
  \href{https://doi.org/10.1103/PhysRevLett.83.4690}{\emph{Phys. Rev. Lett.}
  {\bfseries 83} (1999) 4690}
  [\href{https://arxiv.org/abs/hep-th/9906064}{{\ttfamily hep-th/9906064}}].

\bibitem{Karch:2000ct}
A.~Karch and L.~Randall, \emph{{Locally localized gravity}},
  \href{https://doi.org/10.1088/1126-6708/2001/05/008}{\emph{JHEP} {\bfseries
  05} (2001) 008} [\href{https://arxiv.org/abs/hep-th/0011156}{{\ttfamily
  hep-th/0011156}}].

\bibitem{Karch:2000gx}
A.~Karch and L.~Randall, \emph{{Open and closed string interpretation of SUSY
  CFT's on branes with boundaries}},
  \href{https://doi.org/10.1088/1126-6708/2001/06/063}{\emph{JHEP} {\bfseries
  06} (2001) 063} [\href{https://arxiv.org/abs/hep-th/0105132}{{\ttfamily
  hep-th/0105132}}].

\bibitem{Almheiri:2019hni}
A.~Almheiri, R.~Mahajan, J.~Maldacena and Y.~Zhao, \emph{{The Page curve of
  Hawking radiation from semiclassical geometry}},
  \href{https://doi.org/10.1007/JHEP03(2020)149}{\emph{JHEP} {\bfseries 03}
  (2020) 149} [\href{https://arxiv.org/abs/1908.10996}{{\ttfamily
  1908.10996}}].

\bibitem{Nozaki:2012qd}
M.~Nozaki, T.~Takayanagi and T.~Ugajin, \emph{{Central Charges for BCFTs and
  Holography}}, \href{https://doi.org/10.1007/JHEP06(2012)066}{\emph{JHEP}
  {\bfseries 06} (2012) 066} [\href{https://arxiv.org/abs/1205.1573}{{\ttfamily
  1205.1573}}].

\bibitem{Ugajin:2013xxa}
T.~Ugajin, \emph{{Two dimensional quantum quenches and holography}},
  \href{https://arxiv.org/abs/1311.2562}{{\ttfamily 1311.2562}}.

\bibitem{Miao:2017gyt}
R.-X.~Miao, C.-S.~Chu and W.-Z.~Guo, \emph{{New proposal for a holographic
  boundary conformal field theory}},
  \href{https://doi.org/10.1103/PhysRevD.96.046005}{\emph{Phys. Rev. D}
  {\bfseries 96} (2017) 046005}
  [\href{https://arxiv.org/abs/1701.04275}{{\ttfamily 1701.04275}}].

\bibitem{Chu:2017aab}
C.-S.~Chu, R.-X.~Miao and W.-Z.~Guo, \emph{{On New Proposal for Holographic
  BCFT}}, \href{https://doi.org/10.1007/JHEP04(2017)089}{\emph{JHEP} {\bfseries
  04} (2017) 089} [\href{https://arxiv.org/abs/1701.07202}{{\ttfamily
  1701.07202}}].

\bibitem{Seminara:2017hhh}
D.~Seminara, J.~Sisti and E.~Tonni, \emph{{Corner contributions to holographic
  entanglement entropy in AdS$_{4}$/BCFT$_{3}$}},
  \href{https://doi.org/10.1007/JHEP11(2017)076}{\emph{JHEP} {\bfseries 11}
  (2017) 076} [\href{https://arxiv.org/abs/1708.05080}{{\ttfamily
  1708.05080}}].

\bibitem{Shimaji:2018czt}
T.~Shimaji, T.~Takayanagi and Z.~Wei, \emph{{Holographic Quantum Circuits from
  Splitting/Joining Local Quenches}},
  \href{https://doi.org/10.1007/JHEP03(2019)165}{\emph{JHEP} {\bfseries 03}
  (2019) 165} [\href{https://arxiv.org/abs/1812.01176}{{\ttfamily
  1812.01176}}].

\bibitem{Caputa:2019avh}
P.~Caputa, T.~Numasawa, T.~Shimaji, T.~Takayanagi and Z.~Wei, \emph{{Double
  Local Quenches in 2D CFTs and Gravitational Force}},
  \href{https://doi.org/10.1007/JHEP09(2019)018}{\emph{JHEP} {\bfseries 09}
  (2019) 018} [\href{https://arxiv.org/abs/1905.08265}{{\ttfamily
  1905.08265}}].

\bibitem{Hernandez:2020nem}
J.~Hernandez, R.C.~Myers and S.-M.~Ruan, \emph{{Quantum extremal islands made
  easy. Part III. Complexity on the brane}},
  \href{https://doi.org/10.1007/JHEP02(2021)173}{\emph{JHEP} {\bfseries 02}
  (2021) 173} [\href{https://arxiv.org/abs/2010.16398}{{\ttfamily
  2010.16398}}].

\bibitem{Chalabi:2021jud}
A.~Chalabi, C.P.~Herzog, A.~O'Bannon, B.~Robinson and J.~Sisti, \emph{{Weyl
  anomalies of four dimensional conformal boundaries and defects}},
  \href{https://doi.org/10.1007/JHEP02(2022)166}{\emph{JHEP} {\bfseries 02}
  (2022) 166} [\href{https://arxiv.org/abs/2111.14713}{{\ttfamily
  2111.14713}}].

\bibitem{Kawamoto:2022etl}
T.~Kawamoto, T.~Mori, Y.-k.~Suzuki, T.~Takayanagi and T.~Ugajin,
  \emph{{Holographic local operator quenches in BCFTs}},
  \href{https://doi.org/10.1007/JHEP05(2022)060}{\emph{JHEP} {\bfseries 05}
  (2022) 060} [\href{https://arxiv.org/abs/2203.03851}{{\ttfamily
  2203.03851}}].

\bibitem{Suzuki:2022yru}
Y.-k.~Suzuki and S.~Terashima, \emph{{On the dynamics in the AdS/BCFT
  correspondence}}, \href{https://doi.org/10.1007/JHEP09(2022)103}{\emph{JHEP}
  {\bfseries 09} (2022) 103}
  [\href{https://arxiv.org/abs/2205.10600}{{\ttfamily 2205.10600}}].

\bibitem{Izumi:2022opi}
K.~Izumi, T.~Shiromizu, K.~Suzuki, T.~Takayanagi and N.~Tanahashi, \emph{{Brane
  dynamics of holographic BCFTs}},
  \href{https://doi.org/10.1007/JHEP10(2022)050}{\emph{JHEP} {\bfseries 10}
  (2022) 050} [\href{https://arxiv.org/abs/2205.15500}{{\ttfamily
  2205.15500}}].

\bibitem{Gaberdiel:2001zq}
M.R.~Gaberdiel and A.~Recknagel, \emph{{Conformal boundary states for free
  bosons and fermions}},
  \href{https://doi.org/10.1088/1126-6708/2001/11/016}{\emph{JHEP} {\bfseries
  11} (2001) 016} [\href{https://arxiv.org/abs/hep-th/0108238}{{\ttfamily
  hep-th/0108238}}].

\bibitem{Green:1995ga}
M.B.~Green and M.~Gutperle, \emph{{Symmetry breaking at enhanced symmetry
  points}}, \href{https://doi.org/10.1016/0550-3213(95)00608-7}{\emph{Nucl.
  Phys. B} {\bfseries 460} (1996) 77}
  [\href{https://arxiv.org/abs/hep-th/9509171}{{\ttfamily hep-th/9509171}}].

\bibitem{Recknagel:1998ih}
A.~Recknagel and V.~Schomerus, \emph{{Boundary deformation theory and moduli
  spaces of D-branes}},
  \href{https://doi.org/10.1016/S0550-3213(99)00060-7}{\emph{Nucl. Phys. B}
  {\bfseries 545} (1999) 233}
  [\href{https://arxiv.org/abs/hep-th/9811237}{{\ttfamily hep-th/9811237}}].

\bibitem{Chamblin:2000ra}
A.~Chamblin, H.S.~Reall, H.-a.~Shinkai and T.~Shiromizu, \emph{{Charged brane
  world black holes}},
  \href{https://doi.org/10.1103/PhysRevD.63.064015}{\emph{Phys. Rev. D}
  {\bfseries 63} (2001) 064015}
  [\href{https://arxiv.org/abs/hep-th/0008177}{{\ttfamily hep-th/0008177}}].

\bibitem{Cui:2023gtf}
Z.-Q.~Cui, Y.~Guo and R.-X.~Miao, \emph{{Cone Holography with Neumann Boundary
  Conditions and Brane-localized Gauge Fields}},
  \href{https://arxiv.org/abs/2312.16463}{{\ttfamily 2312.16463}}.

\bibitem{Kaloper:2000xa}
N.~Kaloper, E.~Silverstein and L.~Susskind, \emph{{Gauge symmetry and localized
  gravity in M theory}},
  \href{https://doi.org/10.1088/1126-6708/2001/05/031}{\emph{JHEP} {\bfseries
  05} (2001) 031} [\href{https://arxiv.org/abs/hep-th/0006192}{{\ttfamily
  hep-th/0006192}}].

\bibitem{Lu:2000xc}
H.~Lu and C.N.~Pope, \emph{{Branes on the brane}},
  \href{https://doi.org/10.1016/S0550-3213(01)00021-9}{\emph{Nucl. Phys. B}
  {\bfseries 598} (2001) 492}
  [\href{https://arxiv.org/abs/hep-th/0008050}{{\ttfamily hep-th/0008050}}].

\bibitem{Oda:2000xh}
I.~Oda, \emph{{Reissner-Nordstrom black hole in gravity localized models}},
  \href{https://arxiv.org/abs/hep-th/0008055}{{\ttfamily hep-th/0008055}}.

\bibitem{Herzog:2017xha}
C.P.~Herzog and K.-W.~Huang, \emph{{Boundary Conformal Field Theory and a
  Boundary Central Charge}},
  \href{https://doi.org/10.1007/JHEP10(2017)189}{\emph{JHEP} {\bfseries 10}
  (2017) 189} [\href{https://arxiv.org/abs/1707.06224}{{\ttfamily
  1707.06224}}].

\bibitem{DiPietro:2019hqe}
L.~Di~Pietro, D.~Gaiotto, E.~Lauria and J.~Wu, \emph{{3d Abelian Gauge Theories
  at the Boundary}}, \href{https://doi.org/10.1007/JHEP05(2019)091}{\emph{JHEP}
  {\bfseries 05} (2019) 091}
  [\href{https://arxiv.org/abs/1902.09567}{{\ttfamily 1902.09567}}].

\bibitem{Bartlett-Tisdall:2023ghh}
S.~Bartlett-Tisdall, C.P.~Herzog and V.~Schaub, \emph{{Bootstrapping Boundary
  QED Part I}},  \href{https://arxiv.org/abs/2312.07692}{{\ttfamily
  2312.07692}}.

\bibitem{Dudal:2018mms}
D.~Dudal, A.J.~Mizher and P.~Pais, \emph{{Remarks on the Chern-Simons photon
  term in the QED description of graphene}},
  \href{https://doi.org/10.1103/PhysRevD.98.065008}{\emph{Phys. Rev. D}
  {\bfseries 98} (2018) 065008}
  [\href{https://arxiv.org/abs/1801.08853}{{\ttfamily 1801.08853}}].

\bibitem{Chiodaroli:2011nr}
M.~Chiodaroli, E.~D'Hoker, Y.~Guo and M.~Gutperle, \emph{{Exact half-BPS
  string-junction solutions in six-dimensional supergravity}},
  \href{https://doi.org/10.1007/JHEP12(2011)086}{\emph{JHEP} {\bfseries 12}
  (2011) 086} [\href{https://arxiv.org/abs/1107.1722}{{\ttfamily 1107.1722}}].

\bibitem{Chiodaroli:2012vc}
M.~Chiodaroli, E.~D'Hoker and M.~Gutperle, \emph{{Holographic duals of Boundary
  CFTs}}, \href{https://doi.org/10.1007/JHEP07(2012)177}{\emph{JHEP} {\bfseries
  07} (2012) 177} [\href{https://arxiv.org/abs/1205.5303}{{\ttfamily
  1205.5303}}].

\bibitem{Raamsdonk:2020tin}
M.V.~Raamsdonk and C.~Waddell, \emph{{Holographic and localization calculations
  of boundary F for $ \mathcal{N} $ = 4 SUSY Yang-Mills theory}},
  \href{https://doi.org/10.1007/JHEP02(2021)222}{\emph{JHEP} {\bfseries 02}
  (2021) 222} [\href{https://arxiv.org/abs/2010.14520}{{\ttfamily
  2010.14520}}].

\bibitem{Coccia:2021lpp}
L.~Coccia and C.F.~Uhlemann, \emph{{Mapping out the internal space in AdS/BCFT
  with Wilson loops}},
  \href{https://doi.org/10.1007/JHEP03(2022)127}{\emph{JHEP} {\bfseries 03}
  (2022) 127} [\href{https://arxiv.org/abs/2112.14648}{{\ttfamily
  2112.14648}}].

\bibitem{Sugimoto:2023oul}
S.~Sugimoto and Y.-k.~Suzuki, \emph{{End of the World Branes from Dimensional
  Reduction}},  \href{https://arxiv.org/abs/2312.07891}{{\ttfamily
  2312.07891}}.

\bibitem{Allen:1985wd}
B.~Allen and T.~Jacobson, \emph{{Vector Two Point Functions in Maximally
  Symmetric Spaces}}, \href{https://doi.org/10.1007/BF01211169}{\emph{Commun.
  Math. Phys.} {\bfseries 103} (1986) 669}.

\bibitem{DHoker:1998bqu}
E.~D'Hoker and D.Z.~Freedman, \emph{{Gauge boson exchange in AdS(d+1)}},
  \href{https://doi.org/10.1016/S0550-3213(98)00852-9}{\emph{Nucl. Phys. B}
  {\bfseries 544} (1999) 612}
  [\href{https://arxiv.org/abs/hep-th/9809179}{{\ttfamily hep-th/9809179}}].

\bibitem{DHoker:1999bve}
E.~D'Hoker, D.Z.~Freedman, S.D.~Mathur, A.~Matusis and L.~Rastelli,
  \emph{{Graviton and gauge boson propagators in AdS(d+1)}},
  \href{https://doi.org/10.1016/S0550-3213(99)00524-6}{\emph{Nucl. Phys. B}
  {\bfseries 562} (1999) 330}
  [\href{https://arxiv.org/abs/hep-th/9902042}{{\ttfamily hep-th/9902042}}].

\bibitem{Reeves:2021sab}
W.~Reeves, M.~Rozali, P.~Simidzija, J.~Sully, C.~Waddell and D.~Wakeham,
  \emph{{Looking for (and not finding) a bulk brane}},
  \href{https://doi.org/10.1007/JHEP12(2021)002}{\emph{JHEP} {\bfseries 12}
  (2021) 002} [\href{https://arxiv.org/abs/2108.10345}{{\ttfamily
  2108.10345}}].

\bibitem{Mueck:1998iz}
W.~Mueck and K.S.~Viswanathan, \emph{{Conformal field theory correlators from
  classical field theory on anti-de Sitter space. 2. Vector and spinor
  fields}}, \href{https://doi.org/10.1103/PhysRevD.58.106006}{\emph{Phys. Rev.
  D} {\bfseries 58} (1998) 106006}
  [\href{https://arxiv.org/abs/hep-th/9805145}{{\ttfamily hep-th/9805145}}].

\bibitem{lYi:1998akg}
W.S.~l'Yi, \emph{{Holographic projection of massive vector fields in AdS / CFT
  correspondence}},  \href{https://arxiv.org/abs/hep-th/9808051}{{\ttfamily
  hep-th/9808051}}.

\bibitem{Arutyunov:1998xt}
G.E.~Arutyunov and S.A.~Frolov, \emph{{Antisymmetric tensor field on AdS(5)}},
  \href{https://doi.org/10.1016/S0370-2693(98)01136-8}{\emph{Phys. Lett. B}
  {\bfseries 441} (1998) 173}
  [\href{https://arxiv.org/abs/hep-th/9807046}{{\ttfamily hep-th/9807046}}].

\bibitem{lYi:1998trg}
W.S.~l'Yi, \emph{{Correlators of currents corresponding to the massive p form
  fields in AdS / CFT correspondence}},
  \href{https://doi.org/10.1016/S0370-2693(99)00009-X}{\emph{Phys. Lett. B}
  {\bfseries 448} (1999) 218}
  [\href{https://arxiv.org/abs/hep-th/9811097}{{\ttfamily hep-th/9811097}}].

\bibitem{Suzuki:2022xwv}
K.~Suzuki and T.~Takayanagi, \emph{{BCFT and Islands in two dimensions}},
  \href{https://doi.org/10.1007/JHEP06(2022)095}{\emph{JHEP} {\bfseries 06}
  (2022) 095} [\href{https://arxiv.org/abs/2202.08462}{{\ttfamily
  2202.08462}}].

\bibitem{Kanda:2023zse}
H.~Kanda, M.~Sato, Y.-k.~Suzuki, T.~Takayanagi and Z.~Wei, \emph{{AdS/BCFT with
  brane-localized scalar field}},
  \href{https://doi.org/10.1007/JHEP03(2023)105}{\emph{JHEP} {\bfseries 03}
  (2023) 105} [\href{https://arxiv.org/abs/2302.03895}{{\ttfamily
  2302.03895}}].

\bibitem{Omiya:2021olc}
H.~Omiya and Z.~Wei, \emph{{Causal structures and nonlocality in double
  holography}}, \href{https://doi.org/10.1007/JHEP07(2022)128}{\emph{JHEP}
  {\bfseries 07} (2022) 128}
  [\href{https://arxiv.org/abs/2107.01219}{{\ttfamily 2107.01219}}].

\bibitem{Mori:2023swn}
T.~Mori and B.~Yoshida, \emph{{Exploring causality in braneworld/cutoff
  holography via holographic scattering}},
  \href{https://doi.org/10.1007/JHEP10(2023)104}{\emph{JHEP} {\bfseries 10}
  (2023) 104} [\href{https://arxiv.org/abs/2308.00739}{{\ttfamily
  2308.00739}}].

\bibitem{Liendo:2012hy}
P.~Liendo, L.~Rastelli and B.C.~van Rees, \emph{{The Bootstrap Program for
  Boundary CFT$_d$}},
  \href{https://doi.org/10.1007/JHEP07(2013)113}{\emph{JHEP} {\bfseries 07}
  (2013) 113} [\href{https://arxiv.org/abs/1210.4258}{{\ttfamily 1210.4258}}].

\bibitem{Kaviraj:2018tfd}
A.~Kaviraj and M.F.~Paulos, \emph{{The Functional Bootstrap for Boundary CFT}},
  \href{https://doi.org/10.1007/JHEP04(2020)135}{\emph{JHEP} {\bfseries 04}
  (2020) 135} [\href{https://arxiv.org/abs/1812.04034}{{\ttfamily
  1812.04034}}].

\bibitem{Mazac:2018biw}
D.~Maz\'a\v{c}, L.~Rastelli and X.~Zhou, \emph{{An analytic approach to
  BCFT$_{d}$}}, \href{https://doi.org/10.1007/JHEP12(2019)004}{\emph{JHEP}
  {\bfseries 12} (2019) 004}
  [\href{https://arxiv.org/abs/1812.09314}{{\ttfamily 1812.09314}}].

\bibitem{Kusuki:2021gpt}
Y.~Kusuki, \emph{{Analytic bootstrap in 2D boundary conformal field theory:
  towards braneworld holography}},
  \href{https://doi.org/10.1007/JHEP03(2022)161}{\emph{JHEP} {\bfseries 03}
  (2022) 161} [\href{https://arxiv.org/abs/2112.10984}{{\ttfamily
  2112.10984}}].

\bibitem{Kusuki:2022ozk}
Y.~Kusuki and Z.~Wei, \emph{{AdS/BCFT from conformal bootstrap: construction of
  gravity with branes and particles}},
  \href{https://doi.org/10.1007/JHEP01(2023)108}{\emph{JHEP} {\bfseries 01}
  (2023) 108} [\href{https://arxiv.org/abs/2210.03107}{{\ttfamily
  2210.03107}}].

\bibitem{Gaiotto:2014kfa}
D.~Gaiotto, A.~Kapustin, N.~Seiberg and B.~Willett, \emph{{Generalized Global
  Symmetries}}, \href{https://doi.org/10.1007/JHEP02(2015)172}{\emph{JHEP}
  {\bfseries 02} (2015) 172} [\href{https://arxiv.org/abs/1412.5148}{{\ttfamily
  1412.5148}}].

\bibitem{Gomes:2023ahz}
P.R.S.~Gomes, \emph{{An introduction to higher-form symmetries}},
  \href{https://doi.org/10.21468/SciPostPhysLectNotes.74}{\emph{SciPost Phys.
  Lect. Notes} {\bfseries 74} (2023) 1}
  [\href{https://arxiv.org/abs/2303.01817}{{\ttfamily 2303.01817}}].

\bibitem{Bhardwaj:2023kri}
L.~Bhardwaj, L.E.~Bottini, L.~Fraser-Taliente, L.~Gladden, D.S.W.~Gould,
  A.~Platschorre et~al., \emph{{Lectures on generalized symmetries}},
  \href{https://doi.org/10.1016/j.physrep.2023.11.002}{\emph{Phys. Rept.}
  {\bfseries 1051} (2024) 1}
  [\href{https://arxiv.org/abs/2307.07547}{{\ttfamily 2307.07547}}].

\bibitem{Bergman:2020ifi}
O.~Bergman, Y.~Tachikawa and G.~Zafrir, \emph{{Generalized symmetries and
  holography in ABJM-type theories}},
  \href{https://doi.org/10.1007/JHEP07(2020)077}{\emph{JHEP} {\bfseries 07}
  (2020) 077} [\href{https://arxiv.org/abs/2004.05350}{{\ttfamily
  2004.05350}}].

\bibitem{Hofman:2017vwr}
D.M.~Hofman and N.~Iqbal, \emph{{Generalized global symmetries and
  holography}},
  \href{https://doi.org/10.21468/SciPostPhys.4.1.005}{\emph{SciPost Phys.}
  {\bfseries 4} (2018) 005} [\href{https://arxiv.org/abs/1707.08577}{{\ttfamily
  1707.08577}}].

\end{thebibliography}\endgroup


\end{document}